%% file: main.tex
\documentclass[sigconf,nonacm]{acmart}

\usepackage{subfigure}
\usepackage{siunitx}
\usepackage{enumitem}

\usepackage{algorithm}
\usepackage[noend]{algpseudocode}
\let\ReturnInline\Return
\renewcommand{\Return}{\State\ReturnInline}
\algrenewcommand\algorithmicrequire{$\rhd$}
\algrenewcommand\algorithmicensure{$\square$}

\AtBeginDocument{%
  \providecommand\BibTeX{{%
    \normalfont B\kern-0.5em{\scshape i\kern-0.25em b}\kern-0.8em\TeX}}}

\setcopyright{acmcopyright}
\copyrightyear{2018}
\acmYear{2018}
\acmDOI{XXXXXXX.XXXXXXX}

\acmConference[Conference acronym 'XX]{Make sure to enter the correct
  conference title from your rights confirmation emai}{June 03--05,
  2018}{Woodstock, NY}
\newcommand{\ignore}[1]{}

\begin{document}

%% Full title of the paper.
\title{An Approach for Addressing Internally-Disconnected Communities in Louvain Algorithm}

%% Short title to be used in page headers (optional).
% \title[short title]{full title}
% \subtitle{Something other than the title}

%% Authors and their affiliations.
\author{Subhajit Sahu}
\email{subhajit.sahu@research.iiit.ac.in}
\affiliation{%
  \institution{IIIT Hyderabad}
  \streetaddress{Professor CR Rao Rd, Gachibowli}
  \city{Hyderabad}
  \state{Telangana}
  \country{India}
  \postcode{500032}
}

%% Concise author list in page headers.
%\renewcommand{\shortauthors}{Sahu, Kothapalli, and Banerjee, et al.}

%% Show page numbers.
\settopmatter{printfolios=true}

%% Short summary of the work to be presented in the article.
\begin{abstract}
Community detection is the problem of identifying densely connected clusters within a network. While the Louvain algorithm is commonly used for this task, it can produce internally-disconnected communities. To address this, the Leiden algorithm was introduced. This technical report introduces GSP-Louvain, a parallel algorithm based on Louvain, which mitigates this issue. Running on a system with two 16-core Intel Xeon Gold 6226R processors, GSP-Louvain outperforms Leiden, NetworKit Leiden, and cuGraph Leiden by $391\times$, $6.9\times$, and $2.6\times$ respectively, processing $410M$ edges per second on a $3.8B$ edge graph. Furthermore, GSP-Louvain improves performance at a rate of $1.5\times$ for every doubling of threads.
% Community detection is the problem of identifying densely connected clusters of nodes within a network. The Louvain algorithm is a widely used method for this task, but it can produce communities that are internally disconnected. To address this, the Leiden algorithm was introduced. In this technical report, we propose another approach to mitigate this issue. On a system with two 16-core Intel Xeon Gold 6226R processors, our new parallel algorithm GSP-Louvain, based on the Louvain algorithm, addresses this issue, and outperforms the original Leiden, NetworKit Leiden, and cuGraph Leiden by $391\times$, $6.9\times$, and $2.6\times$ respectively - achieving a processing rate of $410 M$ edges/s on a $3.8 B$ edge graph. Furthermore, GSP-Louvain improves performance at a rate of $1.5\times$ for every doubling of threads.
\end{abstract}

\begin{CCSXML}
<ccs2012>
<concept>
<concept_id>10003752.10003809.10010170</concept_id>
<concept_desc>Theory of computation~Parallel algorithms</concept_desc>
<concept_significance>500</concept_significance>
</concept>
<concept>
<concept_id>10003752.10003809.10003635</concept_id>
<concept_desc>Theory of computation~Graph algorithms analysis</concept_desc>
<concept_significance>500</concept_significance>
</concept>
</ccs2012>
\end{CCSXML}

% \ccsdesc[500]{Theory of computation~Parallel algorithms}
% \ccsdesc[500]{Theory of computation~Graph algorithms analysis}

%% Pick words that accurately describe the work being presented.
\keywords{Community detection, Internally-disconnected communities, Parallel Louvain implementation}


%% Process the author and title information.
\maketitle

\section{Introduction}
\label{sec:introduction}
\input{01-introduction}
\section{Related work}
\label{sec:related}
\input{02-related-work}

\section{Preliminaries}
\label{sec:preliminaries}
\input{03-preliminaries}

\section{Approach}
\label{sec:approach}
\input{04-approach}

\section{Evaluation}
\label{sec:evaluation}
\input{05-evaluation}

\section{Conclusion}
\label{sec:conclusion}
\input{06-conclusion}

%% The acknowledgments section.
\begin{acks}
I would like to thank Prof. Kishore Kothapalli, Prof. Dip Sankar Banerjee, and Vincent Traag for their support.
\end{acks}

%% Bibliography style to be used, and the bibliography file.
\bibliographystyle{ACM-Reference-Format}
\bibliography{main}

\clearpage
\appendix
\section{Appendix}
\input{aa-appendix}

\end{document}

%% file: 01-introduction.tex
Community detection is the process of identifying groups of vertices characterized by dense internal connections and sparse connections between the groups \cite{com-fortunato10}. These groups, referred to as communities or clusters \cite{abbe2018community}, offer valuable insights into the organization and functionality of networks \cite{com-fortunato10}. Community detection finds applications in various fields, including topic discovery in text mining, protein annotation in bioinformatics, recommendation systems in e-commerce, and targeted advertising in digital marketing \cite{com-gregory10}.

Modularity maximization is a frequently employed method for community detection. Modularity measures the difference between the fraction of edges within communities and the expected fraction of edges under random distribution, and ranges from $-0.5$ to $1$\ignore{(higher is better)} \cite{newman2006modularity, com-fortunato10}. However, optimizing for modularity is an NP-hard problem \cite{com-brandes07}. An additional challenge is the lack of prior knowledge about the number and size distribution of communities \cite{com-blondel08}. Heuristic-based approaches are thus used for community detection \cite{clauset2004finding, duch2005community, reichardt2006statistical, com-raghavan07, com-rosvall08, com-blondel08, com-xie11, com-whang13, com-kloster14, com-traag19, com-you20, traag2023large}. The identified communities are considered intrinsic when solely based on network topology and disjoint when each vertex belongs to only one community \cite{com-gregory10, coscia2011classification}.

The Louvain method \cite{com-blondel08} is a popular heuristic-based algorithm for intrinsic and disjoint community detection, and has been identified as one of the fastest and top-performing algorithms \cite{com-lancichinetti09, yang2016comparative}. It utilizes a two-phase approach involving local-moving and aggregation phases to iteratively optimize the modularity metric \cite{newman2006modularity}. Its time complexity is $O(KM)$, where $M$ represents the number of edges in the graph, and $K$ denotes the total number of iterations performed by the algorithm across all passes.

Although widely used, the Louvain method has been noted for generating internally-disconnected and poorly connected communities \cite{com-traag19}. In response, Traag et al. \cite{com-traag19} introduce the Leiden algorithm, which incorporates an extra refinement phase between the local-moving and aggregation phases. This refinement step enables vertices to explore and potentially create sub-communities within the identified communities from the local-moving phase \cite{com-traag19}. However, the proliferation of data and their graph representations has reached unprecedented levels in recent years. In contexts prioritizing scalability, the development of optimized multicore algorithms which address the issue of internally disconnected communities becomes essential, especially in the multicore/shared memory setting\ignore{ --- given its energy efficiency and the prevalence of hardware with large memory capacities}. In this report, we propose GSP-Louvain\footnote{\url{https://github.com/puzzlef/louvain-split-communities-openmp}}, another BFS-based approach for addressing the same issue. This is different from a number of earlier works, which tackled internally-disconnected communities as a post-processing step \cite{com-raghavan07, com-gregory10, hafez2014bnem, luecken2016application, wolf2019paga}.

%% file: 02-related-work.tex
Communities in networks represent functional units or meta-nodes \cite{com-chatterjee19}. Identifying these natural divisions in an unsupervised manner is a crucial graph analytics problem in many domains.\ignore{, including drug discovery, disease prediction, protein annotation, topic discovery, link prediction, recommendation systems, customer segmentation, inferring land use, and criminal identification \cite{com-karatas18}.} Various schemes have been developed for finding communities \cite{clauset2004finding, duch2005community, reichardt2006statistical, com-raghavan07, com-rosvall08, com-blondel08, com-xie11, com-whang13, com-kloster14, com-traag19, com-you20, traag2023large}. These schemes can be categorized into three basic approaches: bottom-up, top-down, and data-structure based, with further classification possible within the bottom-up approach \cite{com-souravlas21}. They can also be classified into divisive, agglomerative, and multi-level methods \cite{com-zarayeneh21}. To evaluate the quality of these methods, fitness scores like modularity are commonly used. Modularity \cite{newman2006modularity} is a metric that ranges from $-0.5$ to $1$, and measures the relative density of links within communities compared to those outside. Optimizing modularity theoretically results in the best possible clustering of the nodes \cite{com-lancichinetti09}. However, as going through each possible clustering of the nodes is impractical \cite{com-fortunato10}, heuristic algorithms such as the Louvain method \cite{com-blondel08} are used.

The Louvain method, proposed by Blondel et al. \cite{com-blondel08} from the University of Louvain, is a greedy, multi-level, modularity-optimization based algorithm for intrinsic and disjoint community detection \cite{com-lancichinetti09, newman2006modularity}. It utilizes a two-phase approach involving local-moving and aggregation phases to iteratively optimize the modularity metric \cite{com-blondel08}, and is recognized as one of the fastest and top-performing algorithms \cite{com-lancichinetti09, yang2016comparative}. Various algorithmic improvements \cite{com-ryu16, com-halappanavar17, com-lu15, com-waltman13, com-rotta11, com-shi21} and parallelization strategies \cite{com-lu15, com-halappanavar17, com-fazlali17, com-wickramaarachchi14, com-cheong13, com-zeng15} have been proposed for the Louvain method. Several open-source implementations and software packages have been developed for community detection using Parallel Louvain Algorithm, including Vite \cite{ghosh2018scalable}, Grappolo \cite{com-halappanavar17}, and NetworKit \cite{staudt2016networkit}.\ignore{Note however that community detection methods which rely on modularity maximization are known to face the resolution limit problem, which hinders the identification of smaller communities \cite{fortunato2007resolution}.}

Although favored for its ability to identify communities with high modularity, the Louvain method often produces internally disconnected communities due to vertices acting as bridges moving to other communities during iterations (see Section \ref{sec:about-louvain-disconnected}). This issue worsens with further iterations without decreasing the quality function \cite{com-traag19}. To overcome these issues, Traag et al. \cite{com-traag19} from the University of Leiden, propose the Leiden algorithm, which introduces a refinement phase after the local-moving phase of the Louvain method. In this refinement phase, vertices undergo additional local moves based on delta-modularity, allowing the discovery of sub-communities within the initial communities. Shi et al. \cite{com-shi21} also introduce a similar refinement phase after the local-moving phase of the Louvain method to minimize poor clusters.

A few open-source implementations and software packages exist for community detection using the Leiden algorithm. The original implementation, \texttt{libleidenalg} \cite{com-traag19}, is written in C++ and has a Python interface called \texttt{leidenalg}. NetworKit \cite{staudt2016networkit}, a software package designed for analyzing graph data sets with billions of connections, features a parallel implementation of the Leiden algorithm by Nguyen \cite{nguyenleiden}. This implementation utilizes global queues for vertex pruning and vertex and community locking for updating communities. Another option, cuGraph \cite{kang2023cugraph}, is a GPU-accelerated library for graph analytics, part of the RAPIDS suite. By leveraging NVIDIA GPUs, cuGraph provides significant performance improvements over traditional CPU-based approaches. Its core is written in C++ using CUDA, with a Python interface designed to offer an accessible experience for data scientists and developers.

Internally disconnected communities are not a new problem. Internally-disconnected communities can be produced by label propagation algorithms \cite{com-raghavan07, com-gregory10}, multi-level algorithms \cite{com-blondel08}, genetic algorithms \cite{hesamipour2022detecting}, expectation minimization/maximization algorithms \cite{ball2011efficient, hafez2014bnem}, among others. Raghavan et al. \cite{com-raghavan07} note that their Label Propagation Algorithm (LPA) for community detection may identify disconnected communities. In such a case, they suggest applying a Breadth-First Search (BFS) on the subnetworks of each individual group to separate the disconnected communities, with a time complexity of $O(M+N)$. Gregory \cite{com-gregory10} introduced the Community Overlap PRopagation Algorithm (COPRA) as an extension of LPA. In their algorithm, they eliminate communities completely contained within others and use a similar method as Raghavan et al. \cite{com-raghavan07} to split any returned disconnected communities into connected ones. Hafez et al. \cite{hafez2014bnem} propose a community detection algorithm using Bayesian Network and Expectation Maximization (BNEM). To address cases where a single community contains disconnected sub-communities, they relabel these components, which occurs when the network has more communities than the specified number $k$. Hesamipour et al. \cite{hesamipour2022detecting} introduce a genetic algorithm that combines similarity-based and modularity-based methods, using a Minimum Spanning Tree (MST) representation to handle disconnected communities and improve mutation effectiveness.
% Hafez et al. \cite{hafez2014bnem} present a community detection algorithm leveraging Bayesian Network and Expectation Minimization (BNEM). They include a final step to examine the result for potentially containing disconnected communities within a single community. If this situation arises, new community labels are assigned to the disconnected components. This scenario occurs when the network may have more communities than the specified number $k$ in the algorithm. Hesamipour et al. \cite{hesamipour2022detecting} propose a genetic algorithm for community detection, integrating similarity-based and modularity-based approaches. Their method uses an MST-based representation for addressing issues like disconnected communities and ineffective mutations.

\ignore{Goel et al. \cite{goel2023effective} present a tangential work, where they identify a set of nodes in a network, known as a Structural Hole Spanner (SHS), that act as a bridge among different otherwise disconnected communities.}

%% file: 03-preliminaries.tex
Consider an undirected graph denoted as $G(V, E, w)$, where $V$ stands for the vertex set, $E$ represents the edge set, and $w_{ij} = w_{ji}$ indicates the weight associated with each edge. In the scenario of an unweighted graph, we assume a unit weight for every edge ($w_{ij} = 1$). Moreover, the neighbors of a vertex $i$ are referred to as $J_i = \{j\ |\ (i, j) \in E\}$, the weighted degree of each vertex as $K_i = \sum_{j \in J_i} w_{ij}$, the total number of vertices as $N = |V|$, the total number of edges as $M = |E|$, and the sum of edge weights in the undirected graph as $m = \sum_{i, j \in V} w_{ij}/2$.

\subsection{Community detection}

Disjoint community detection entails identifying a community membership mapping, $C: V \rightarrow \Gamma$, wherein each vertex $i \in V$ is assigned a community ID $c$ from the set of community IDs $\Gamma$. We denote the vertices of a community $c \in \Gamma$ as $V_c$, and the community that a vertex $i$ belongs to as $C_i$. Additionally, we denote the neighbors of vertex $i$ belonging to a community $c$ as $J_{i \rightarrow c} = \{j\ |\ j \in J_i\ and\ C_j = c\}$, the sum of those edge weights as $K_{i \rightarrow c} = \sum_{j \in J_{i \rightarrow c}} w_{ij}$, the sum of weights of edges within a community $c$ as $\sigma_c = \sum_{(i, j) \in E\ and\ C_i = C_j = c} w_{ij}$, and the total edge weight of a community $c$ as $\Sigma_c = \sum_{(i, j) \in E\ and\ C_i = c} w_{ij}$ \cite{com-leskovec21}.

\subsection{Modularity}

Modularity is a metric used for assessing the quality of communities identified by heuristic-based community detection algorithms. It is calculated as the difference between the fraction of edges within communities and the expected fraction if edges were randomly distributed, and ranges from $-0.5$ to $1$, where higher values indicate superior results \cite{com-brandes07}.\ignore{The optimization of this metric theoretically leads to the optimal grouping \cite{com-newman04, com-traag11}.} The modularity $Q$ of identified communities is determined using Equation \ref{eq:modularity}, where $\delta$ represents the Kronecker delta function ($\delta (x,y)=1$ if $x=y$, $0$ otherwise). The \textit{delta modularity} of moving a vertex $i$ from community $d$ to community $c$, denoted as $\Delta Q_{i: d \rightarrow c}$, can be determined using Equation \ref{eq:delta-modularity}.

\begin{equation}
\label{eq:modularity}
  Q
  = \frac{1}{2m} \sum_{(i, j) \in E} \left[w_{ij} - \frac{K_i K_j}{2m}\right] \delta(C_i, C_j)
  = \sum_{c \in \Gamma} \left[\frac{\sigma_c}{2m} - \left(\frac{\Sigma_c}{2m}\right)^2\right]
\end{equation}

\begin{equation}
\label{eq:delta-modularity}
  \Delta Q_{i: d \rightarrow c}
  = \frac{1}{m} (K_{i \rightarrow c} - K_{i \rightarrow d}) - \frac{K_i}{2m^2} (K_i + \Sigma_c - \Sigma_d)
\end{equation}

\subsection{Louvain algorithm}
\label{sec:about-louvain}

The Louvain method, introduced by Blondel et al. \cite{com-blondel08}, is a greedy algorithm that optimizes modularity to identify high quality disjoint communities in large networks. It has a time complexity of $O (L |E|)$, where $L$ is the total number of iterations performed, and a space complexity of $O(|V| + |E|)$ \cite{com-lancichinetti09}. This algorithm consists of two phases: the \textit{local-moving phase}, wherein each vertex $i$ greedily decides to join the community of one of its neighbors $j \in J_i$ to maximize the gain in modularity $\Delta Q_{i:C_i \rightarrow C_j}$ (using Equation \ref{eq:delta-modularity}), and the \textit{aggregation phase}, during which all vertices within a community are combined into a single super-vertex. These phases constitute one pass, which is repeated until no further improvement in modularity is achieved \cite{com-blondel08, com-leskovec21}.

\subsection{Possibility of Internally-disconnected communities with the Louvain algorithm}
\label{sec:about-louvain-disconnected}

\input{src/fig-onlouvain}

The Louvain method, though effective, has been noted to potentially identify internally disconnected communities \cite{com-traag19}. This is illustrated with an example in Figure \ref{fig:onlouvain}. Figure \ref{fig:onlouvain--1} shows the initial community structure after running a few iterations of the Louvain algorithm. It includes four communities labeled $C1$, $C2$, $C3$, $C4$, and vertices $1$ to $7$ are grouped in community $C1$\ignore{, with thick lines indicating higher edge weights}. After a few additional iterations, in Figure \ref{fig:onlouvain--2}, communities $C2$, $C3$, and $C4$ merge together into $C3$, due to strong connections among themselves. As vertex $4$ is now more strongly connected to community $C3$, it shifts from community $C1$ to join community $C3$ instead, in Figure \ref{fig:onlouvain--3}. This results in the internal disconnection of community $C1$, as vertices $1$, $2$, $3$, $5$, $6$, and $7$ retain their locally optimal assignments. Additionally, once all nodes are optimally assigned, the algorithm aggregates the graph. If an internally disconnected community becomes a node in the aggregated graph, it remains disconnected unless it combines with another community acting as a bridge. With subsequent passes, these disconnected communities are prone to steering the solution towards a lower local optima.

\subsection{Leiden algorithm}
\label{sec:about-leiden}

The Leiden algorithm, proposed by Traag et al. \cite{com-traag19}, is a multi-level community detection technique that extends the Louvain method. It consists of three key phases.\ignore{: the local-moving phase, the refinement phase, and the aggregation phase.} In the \textit{local-moving phase}, each vertex $i$ optimizes its community assignment by greedily selecting to join the community of one of its neighbors $j \in J_i$, aiming to maximize the modularity gain $\Delta Q_{i:C_i \rightarrow C_j}$, as defined by Equation \ref{eq:delta-modularity}, akin to the Louvain method. During the \textit{refinement phase}, vertices within each community undergo further updates to their community memberships, starting from singleton partitions. Unlike the local-moving phase however, these updates are not strictly greedy. Instead, vertices may move to any community within their bounds where the modularity increases, with the probability of joining a neighboring community proportional to the delta-modularity of the move\ignore{, as computed by Equation \ref{eq:delta-modularity}}. The level of randomness in these moves is governed by a parameter $\theta > 0$. This randomized approach facilitates the identification of higher quality sub-communities within the communities established during the local-moving phase. Finally, in the \textit{aggregation phase}, all vertices within each refined partition are combined into super-vertices, with an initial community assignment derived from the local-moving phase \cite{com-traag19}. The time complexity of the Leiden algorithm is $O(L|E|)$, where $L$ represents the total number of iterations performed, and its space complexity is $O(|V| + |E|)$\ignore{ --- similar to the Louvain method}.

%% file: src/fig-onlouvain.tex
\begin{figure*}[hbtp]
  \centering
  \subfigure[Initial community structure with $4$ communities]{
    \label{fig:onlouvain--1}
    \includegraphics[width=0.31\linewidth]{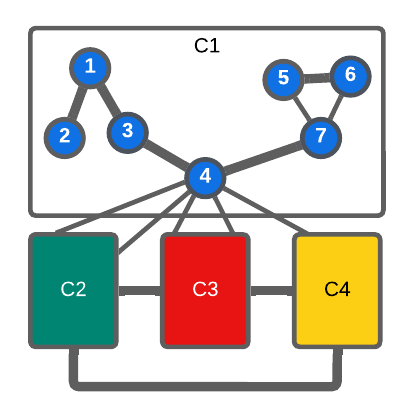}
  }
  \subfigure[After few iterations, $C2$, $C3$, and $C4$ merge]{
    \label{fig:onlouvain--2}
    \includegraphics[width=0.31\linewidth]{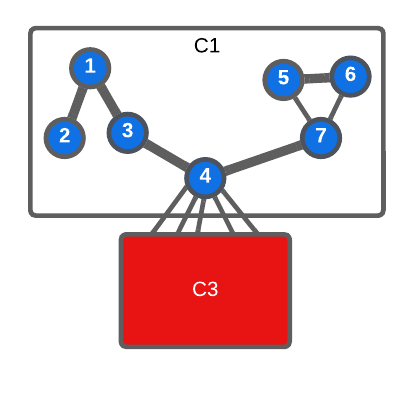}
  }
  \subfigure[Subsequently, vertex $4$ moves to $C3$]{
    \label{fig:onlouvain--3}
    \includegraphics[width=0.31\linewidth]{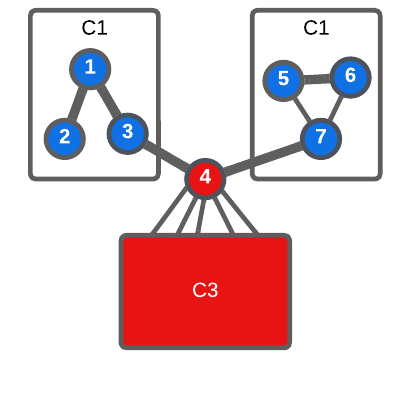}
  } \\[-2ex]
  \caption{An example demonstrating the possibility of internally disconnected communities with the Louvain algorithm. Here, $C1$, $C2$, $C3$, and $C4$ are four communities obtained after running a few iterations of the Louvain algorithm, with vertices $1$ to $7$ being members of community $C1$. Thick lines are used to denote higher edge weights.}
  \label{fig:onlouvain}
\end{figure*}

%% file: 04-approach.tex
In preceding sections, we explored how the Louvain algorithm can produce internally-disconnected communities. However, this phenomenon is not exclusive to this algorithms and has been documented in various other community detection algorithms \cite{com-raghavan07, com-gregory10, hesamipour2022detecting, ball2011efficient, hafez2014bnem}. To mitigate this issue, a commonly employed approach involves splitting disconnected communities as a post-processing step \cite{com-raghavan07, com-gregory10, hafez2014bnem, luecken2016application, wolf2019paga}, utilizing Breadth First Search (BFS) \cite{com-raghavan07, com-gregory10}. We refer to this as Split Last (SL) using BFS, or \textit{SL-BFS}. However, this strategy may exacerbate the problem of poorly connected communities for multi-level community detection techniques, such as the Louvain algorithm \cite{com-traag19}.

\subsection{Our Split Pass (SP) approach}

To tackle the aforementioned challenges encountered by the Louvain algorithm, we propose to split the disconnected communities in every pass. Specifically, this occurs after the local-moving phase in the Louvain algorithm. We refer to this as the \textit{Split Pass (SP)} approach. Additionally, we explore the conventional approach of splitting disconnected communities as a post-processing step, i.e., after all iterations of the community detection algorithm have been completed and the vertex community memberships have converged. This traditional approach is referred to as \textit{Split Last (SL)}.

In order to partition disconnected communities using either the SP or the SL approach, we explore three distinct techniques: minimum-label-based \textit{Label Propagation (LP)}, minimum-label-based \textit{Label Propagation with Pruning (LPP)}, and \textit{Breadth First Search (BFS)}. The rationale behind investigating LP and LPP techniques for splitting disconnected communities as they are readily parallelizable.

With the LP technique, each vertex in the graph initially receives a unique label (its vertex ID). Subsequently, in each iteration, every vertex selects the minimum label among its neighbors within its assigned community, as determined by the community detection algorithm. This iterative process continues until labels for all vertices converge. Since each vertex obtains a unique label within its connected component and its community, communities comprising multiple connected components get partitioned. In contrast, the LPP technique incorporates a pruning optimization step where only unprocessed vertices are handled. Once a vertex is processed, it is marked as such, and gets reactivated (or marked as unprocessed) if one of its neighbors changes their label. The pseudocode for the LP and LPP techniques is presented in Algorithm \ref{alg:splitlp}, with detailed explanations given in Section \ref{sec:splitlp}.

On the other hand, the BFS technique for splitting internally-disconnected communities involves selecting a random vertex within each community and identifying all vertices reachable from it as part of one subcommunity. If any vertices remain unvisited in the original community, another random vertex is chosen from the remaining set, and the process iterates until all vertices within each community are visited. Consequently, the BFS technique facilitates the partitioning of connected components within each community. The pseudocode for the BFS technique for splitting disconnected communities is outlined in Algorithm \ref{alg:splitbfs}, with its in-depth explanation provided in Section \ref{sec:splitbfs}.

\subsubsection{Explanation of LP/LPP algorithm\ignore{ for splitting disconnected communities}}
\label{sec:splitlp}

We now discuss the pseudocode for the parallel minimum-label-based Label Propagation (LP) and Label Propagation with Pruning (LPP) techniques, given in Algorithm \ref{alg:splitlp}, that partition the internally-disconnected communities. These techniques can be employed either as a post-processing step (SL) at the end of the community detection algorithm, or after the refinement/local-moving phase (SP) in each pass. Here, the function \texttt{splitDisconnectedLp()} that is responsible for this task, takes as input the graph $G(V, E)$ and the community memberships $C$ of vertices, and returns the updated community memberships $C'$ where all disconnected communities have been separated.

In lines \ref{alg:splitlp--init-begin}-\ref{alg:splitlp--init-end}, the algorithm starts by initializing the minimum labels $C'$ of each vertex to their respective vertex IDs, and designates all vertices as unprocessed. Lines \ref{alg:splitlp--loop-begin}-\ref{alg:splitlp--loop-end} represent the iteration loop of the algorithm. Initially, the number of changed labels $\Delta N$ is initialized (line \ref{alg:splitlp--loopinit}). This is followed by an iteration of label propagation (lines \ref{alg:splitlp--vertexloop-begin}-\ref{alg:splitlp--vertexloop-end}), and finally a convergence check (line \ref{alg:splitlp--checkconverged}). During each iteration (lines \ref{alg:splitlp--vertexloop-begin}-\ref{alg:splitlp--vertexloop-end}), unprocessed vertices are processed in parallel. For each unprocessed vertex $i$, it is marked as processed if the Label Propagation with Pruning (LPP) technique is utilized. The algorithm proceeds to identify the minimum label $c'_{min}$ within the community of vertex $i$ (lines \ref{alg:splitlp--findmin-begin}-\ref{alg:splitlp--findmin-end}). This is achieved by iterating over the outgoing neighbors of $i$ in the graph $G$ and considering only those neighbors belonging to the same community as $i$. The minimum label found among these neighbors, along with the label of vertex $i$ itself, determines the minimum community label $c'_{min}$ for $i$. If the obtained minimum label $c'_{min}$ differs from the current minimum label $C'[i]$ (line \ref{alg:splitlp--skipmin}), $C'[i]$ is updated, $\Delta N$ is incremented to reflect the change, and neighboring vertices of $i$ belonging to the same community as $i$ are marked as unprocessed to facilitate their reassessment in subsequent iterations. The label propagation loop (lines \ref{alg:splitlp--loop-begin}-\ref{alg:splitlp--loop-end}) continues until there are no further changes in the minimum labels. Finally, the updated labels $C'$, representing the updated community membership of each vertex with no disconnected communities, are returned in line \ref{alg:splitlp--return}.

\input{src/alg-splitlp}

\subsubsection{Explanation of BFS algorithm\ignore{ for splitting disconnected communities}}
\label{sec:splitbfs}

Next, we proceed to describe the pseudocode of the parallel Breadth First Search (BFS) technique, as presented in Algorithm \ref{alg:splitbfs}, devised for the partitioning of disconnected communities. As with LP/LPP techniques, this technique can be applied either as a post-processing step (SL) at the end or after the refinement or local-moving phase (SP) in each pass. Here, the function \texttt{splitDisconnectedBfs()} accepts the input graph $G(V, E)$ and the community membership $C$ of each vertex, and returns the updated community membership $C'$ of each vertex where all the disconnected communities have been split.

\input{src/alg-splitbfs}

Initially, in lines \ref{alg:splitbfs--init-begin}-\ref{alg:splitbfs--init-end}, the flag vector $vis$ representing visited vertices is initialized, the flag vector $busy$ indicating whether a community is currently being processed by a thread, and the labels $C'$ for each vertex are set to their corresponding vertex IDs. Subsequently, each thread concurrently processes every vertex $i$ in the graph $G$ (lines \ref{alg:splitbfs--loop-begin}-\ref{alg:splitbfs--loop-end}). If vertex $i$ has been visited, or the community $c$ of vertex $i$ is \textit{busy}, i.e., it is being processed by some other thread, the current thread proceeds to the next iteration (line \ref{alg:splitbfs--work}). Conversely, if vertex $i$ has not been visited, and the community $c$ of vertex $i$ is not busy, the current thread attempts to mark community $c$ as \textit{busy} using $atomicCAS()$ (line \ref{alg:splitbfs--atomiccas}). If the current thread fails to successfully mark community $c$ as busy, it moves on to the next iteration. If, however, community $c$ was successfully marked as busy, a BFS is performed from vertex $i$ to explore vertices within the same community. This BFS utilizes lambda functions $f_{if}$ to selectively execute BFS on vertex $j$ if it belongs to the same community, and $f_{do}$ to update the label of visited vertices after each vertex is explored during BFS (line \ref{alg:splitbfs--bfs}). Upon completion of processing all vertices, threads synchronize, and the revised labels $C'$ --- representing the updated community membership of each vertex with no disconnected communities --- are returned (line \ref{alg:splitbfs--return}). Finally, the community $c$ is marked as \textit{not busy}, allowing it to be processed by other threads.

Figure \ref{fig:onsplitbfs} illustrates an example of the BFS technique. Initially, Figure \ref{fig:onsplitbfs--1} displays two communities, $C1$ and $C2$, derived after the local-moving phase. Here, $C1$ has become internally disconnected due to the inclusion of vertex $4$ into community $C2$ --- similar to the case depicted in Figure \ref{fig:onlouvain--3}. Subsequently, employing the BFS technique, a thread selects a random vertex within community $C1$, such as $2$, and designates all vertices reachable within $C1$ from $2$ with the label of $2$, and marking them as visited (Figure \ref{fig:onsplitbfs--2}). Following this, as depicted in Figure \ref{fig:onsplitbfs--3}, another thread picks an unvisited vertex randomly within community $C1$, for example, $7$, and labels all vertices reachable within $C1$ from $7$ with the label of $7$, and marking them as visited. An analogous process is executed within community $C2$. Consequently, all vertices are visited, and the labels assigned to them denote the updated community membership of each vertex with no disconnected communities.
% Note that each thread has a mutually exclusive work-list, ensuring that two threads do not simultaneously perform BFS within the same community.

\input{src/fig-onsplitbfs}

\subsection{Our GSP-Louvain algorithm}

To assess both our Split Pass (SP) approach and the conventional Split Last (SL) approach, utilizing minimum-label-based Label Propagation (LP), minimum-label-based Label Propagation with Pruning (LPP), and Breadth First Search (BFS) techniques for splitting disconnected communities with the Louvain algorithm, we use GVE-Louvain \cite{sahu2023gvelouvain}, our parallel implementation of Louvain algorithm.

\subsubsection{Determining suitable technique for splitting disconnected communities}

We now determine the optimal technique for partitioning internally-disconnected communities using GVE-Louvain. To achieve this, we investigate both the SL and SP approaches, employing LP, LPP, and BFS techniques. Figures \ref{fig:optlousp--runtime}, \ref{fig:optlousp--modularity}, and \ref{fig:optlousp--disconnected} illustrate the mean relative runtime, modularity, and fractions of disconnected communities for SL-LP, SL-LPP, SL-BFS, SP-LP, SP-LPP, SP-BFS, and the default (i.e., not splitting disconnected communities) approaches. As depicted in Figure \ref{fig:optlousp--disconnected}, both SL and SP approaches result in non-disconnected communities. Additionally, Figure \ref{fig:optlousp--modularity} reveals that the modularity of communities obtained through the SP approach surpasses that of the SL approach while maintaining proximity to the default approach. Finally, Figure \ref{fig:optlousp--runtime} illustrates that SP-BFS, specifically the SP approach employing the BFS technique, demonstrates superior performance. Consequently, employing BFS to split disconnected communities in each pass (SP) of the Louvain algorithm emerges as the preferred choice.

\input{src/fig-optlousp}

\subsubsection{Explanation of the algorithm}

We refer to GVE-Louvain, that employs the Split Pass approach with Breadth-First Search (SP-BFS) to handle disconnected communities, as \textit{GSP-Louvain}. The core procedure of GSP-Louvain, \texttt{louvain()}, is given in Algorithm \ref{alg:louvain}. It consists of the following main steps: initialization, local-moving phase, splitting phase, and the aggregation phase. The function takes as input a graph $G(V, E)$ and outputs the community membership $C$ for each vertex in the graph, with none of the returned communities being internally disconnected.

First, in line \ref{alg:louvain--initialization}, the community membership $C$ is initialized for each vertex in $G$, and the algorithm conducts passes of the Louvain algorithm, limited to a maximum number of passes defined by $MAX\_PASSES$. During each pass, various metrics such as the total edge weight of each vertex $K'$, the total edge weight of each community $\Sigma'$, and the community membership $C'$ of each vertex in the current graph $G'$ are updated. Subsequently, in line \ref{alg:louvain--local-move}, the local-moving phase is executed by invoking \texttt{louvainMove()} (Algorithm \ref{alg:commonlm}), which optimizes the community assignments. Following this, the algorithm proceeds to the splitting phase, where the internally disconnected communities in $C'$ are separated. This is done using the parallel BFS technique, in line \ref{alg:louvain--split}, with the \texttt{splitCommunitiesBfs()} function (Algorithm \ref{alg:splitbfs}). Next, in line \ref{alg:louvain--globally-converged}, global convergence is inferred if the local-moving phase converges in a single iteration. If so, we terminate the passes. Additionally, if there is minimal reduction in the number of communities ($|\Gamma|$), indicating diminishing returns, the current pass is halted (line \ref{alg:louvain--aggregation-tolerance}).

If convergence is not achieved, the algorithm proceeds with the following steps: renumbering communities (line \ref{alg:louvain--renumber}), updating top-level community memberships $C$ using dendrogram lookup (line \ref{alg:louvain--lookup}), executing the aggregation phase via \texttt{louvainAggregate()} (Algorithm \ref{alg:commonag}), and adjusting the convergence threshold for subsequent passes, known as threshold scaling (line \ref{alg:louvain--threshold-scaling}). The subsequent pass initiates at line \ref{alg:louvain--passes-begin}. Upon completion of all passes, a final update of the top-level community memberships $C$ using dendrogram lookup occurs (line \ref{alg:louvain--lookup-last}), followed by the return of the top-level community membership $C$ of each vertex in graph $G$.

\input{src/alg-louvain}

%% file: src/alg-splitlp.tex
\begin{algorithm}[hbtp]
\caption{Split disconnected communities using (min) LP.}
\label{alg:splitlp}
\begin{algorithmic}[1]
\Require{$G(V, E)$: Input graph}
\Require{$C$: Initial community membership/label of each vertex}
\Ensure{$C'$: Updated community membership/label of each vertex}
\Ensure{$c'_{min}$: Minimum connected label within the community}
\Ensure{$\Delta N$: Number of changes in labels}

\Statex

\Function{splitDisconnectedLp}{$G, C$} \label{alg:splitlp--begin}
  \State $C' \gets \{\}$ \label{alg:splitlp--init-begin}
  \ForAll{$i \in V$ \textbf{in parallel}}
    \State Mark $i$ as unprocessed
    \State $C'[i] = i$
  \EndFor \label{alg:splitlp--init-end}
  \Loop \label{alg:splitlp--loop-begin}
    \State $\Delta N \gets 0$ \label{alg:splitlp--loopinit}
    \ForAll{unprocessed $i \in V$ \textbf{in parallel}} \label{alg:splitlp--vertexloop-begin}
      \If{\textbf{is SL-LPP or SP-LPP}}
        \State Mark $i$ as processed
      \EndIf
      \State $\rhd$ Find minimum community label
      \State $c'_{min} \gets C'[i]$ \label{alg:splitlp--findmin-begin}
      \ForAll{$j \in G.out(i)$}
        \If{$C[j] = C[i]$}
          \State $c'_{min} \gets min(C'[j], c'_{min})$
        \EndIf
      \EndFor \label{alg:splitlp--findmin-end}
      \If{$c'_{min} = C'[i]$} \textbf{continue} \label{alg:splitlp--skipmin}
      \EndIf
      \State $\rhd$ Update community label
      \State $C'[i] \gets c'_{min}$ \textbf{;} $\Delta N \gets \Delta N + 1$ \label{alg:splitlp--updatemin--begin}
      \If{\textbf{is SL-LPP or SP-LPP}}
        \ForAll{$j \in G.out(i)$}
          \State Mark $j$ as unprocessed \textbf{if} $C[j] = C[i]$
        \EndFor
      \EndIf \label{alg:splitlp--updatemin--end}
    \EndFor \label{alg:splitlp--vertexloop-end}
    \State $\rhd$ Converged?
    \If{$\Delta N = 0$} \textbf{break} \label{alg:splitlp--checkconverged}
    \EndIf
  \EndLoop \label{alg:splitlp--loop-end}
  \Return{$C'$} \label{alg:splitlp--return}
\EndFunction \label{alg:splitlp--end}
\end{algorithmic}
\end{algorithm}

%% file: src/alg-splitbfs.tex
\begin{algorithm}[hbtp]
\caption{Split disconnected communities using BFS.}
\label{alg:splitbfs}
\begin{algorithmic}[1]
\Require{$G(V, E)$: Input graph}
\Require{$C$: Initial community membership/label of each vertex}
\Ensure{$C'$: Updated community membership/label of each vertex}
\Ensure{$f_{if}$: Perform BFS to vertex $j$ if condition satisfied}
\Ensure{$f_{do}$: Perform operation after each vertex is visited}
\Ensure{$busy$: Is a community being processed by a thread?}
\Ensure{$vis$: Visited flag for each vertex}

\Statex

\Function{splitDisconnectedBfs}{$G, C$} \label{alg:splitbfs--begin}
  \State $C' \gets vis \gets busy \gets \{\}$ \label{alg:splitbfs--init-begin}
  \ForAll{$i \in V$ \textbf{in parallel}}
    \State $C'[i] = i$
  \EndFor \label{alg:splitbfs--init-end}
  \ForAll{\textbf{threads}} \label{alg:splitbfs--threads-begin}
    \ForAll{$i \in V$} \label{alg:splitbfs--loop-begin}
      \State $c \gets C[i]$ \textbf{;} $c' \gets C'[i]$ \label{alg:splitbfs--loopinit}
      \If{$vis[i]$ \textbf{or} $busy[c]$} \textbf{continue} \label{alg:splitbfs--work}
      \EndIf
      \If{$atomicCAS(busy[c], 0, 1)$ \textbf{fails}} \textbf{continue} \label{alg:splitbfs--atomiccas}
      \EndIf
      \State $f_{if} \gets (j) \implies C[j] = C[j]$
      \State $f_{do} \gets (j) \implies C'[j] \gets c'$
      \State $bfsVisitForEach(vis, G, i, f_{if}, f_{do})$ \label{alg:splitbfs--bfs}
      \State $busy[c] \gets 0$
    \EndFor \label{alg:splitbfs--loop-end}
  \EndFor \label{alg:splitbfs--threads-end}
  \Return{$C'$} \label{alg:splitbfs--return}
\EndFunction \label{alg:splitbfs--end}
\end{algorithmic}
\end{algorithm}

%% file: src/fig-onsplitbfs.tex
\begin{figure*}[hbtp]
  \centering
  \subfigure[Community $C1$ is internally disconnected]{
    \label{fig:onsplitbfs--1}
    \includegraphics[width=0.28\linewidth]{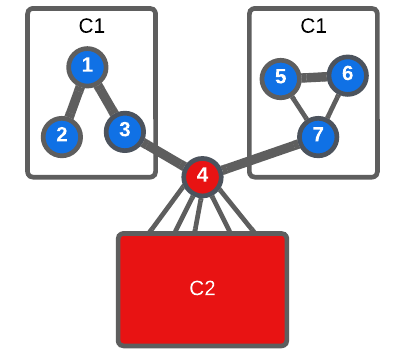}
  }
  \subfigure[After BFS relabeling from vertex $2$]{
    \label{fig:onsplitbfs--2}
    \includegraphics[width=0.28\linewidth]{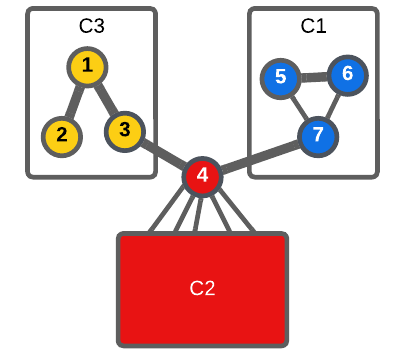}
  }
  \subfigure[After BFS relabeling from vertex $7$]{
    \label{fig:onsplitbfs--3}
    \includegraphics[width=0.28\linewidth]{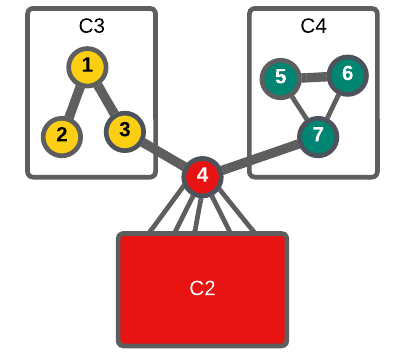}
  } \\[-2ex]
  \caption{An example illustrating the BFS technique for splitting internally-disconnected communities. Initially, two communities, $C1$ and $C2$, are shown, with $C1$ being internally disconnected due to vertex $4$ joining $C2$. The BFS technique selects random vertices within each community and labels reachable vertices with the same label, indicated with a new community ID.}
  \label{fig:onsplitbfs}
\end{figure*}

%% file: src/fig-optlousp.tex
\begin{figure}[hbtp]
  \centering
  \subfigure[Relative runtime using different approaches for splitting disconnected communities with Parallel Louvain algorithm]{
    \label{fig:optlousp--runtime}
    \includegraphics[width=0.98\linewidth]{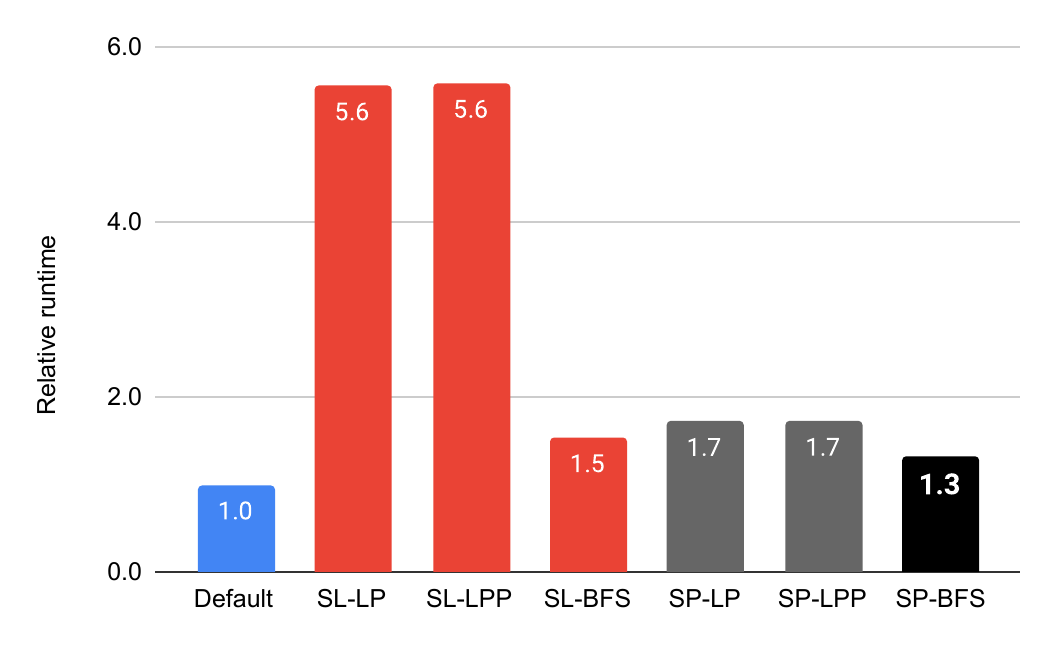}
  } \\[-0ex]
  \subfigure[Modularity using different approaches for splitting disconnected communities with Parallel Louvain algorithm]{
    \label{fig:optlousp--modularity}
    \includegraphics[width=0.98\linewidth]{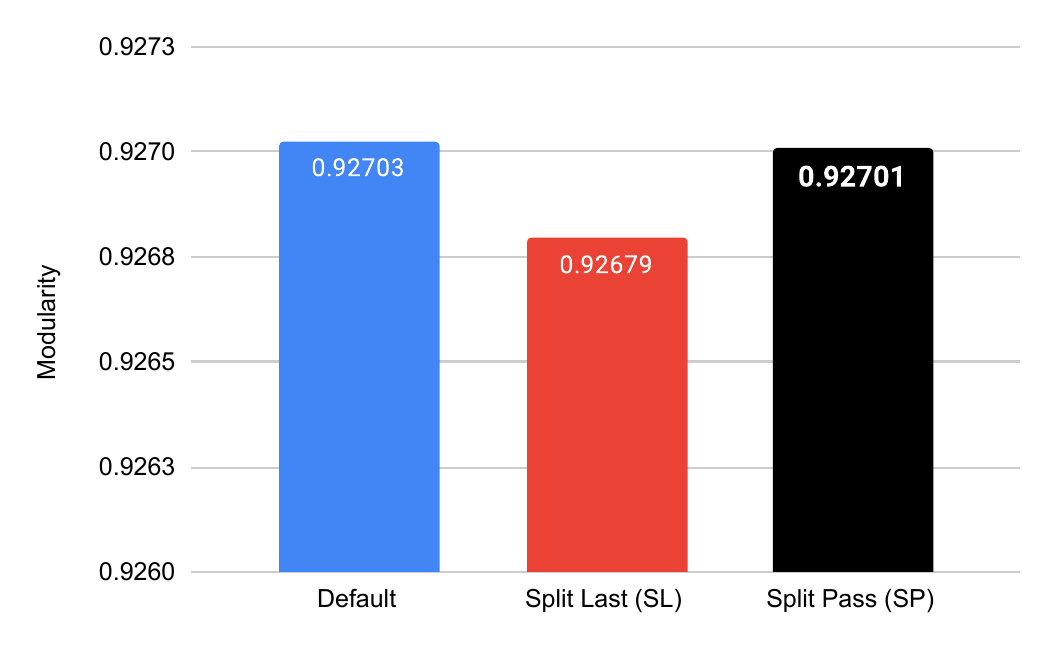}
  } \\[-0ex]
  \subfigure[Fraction of disconnected communities (logarithmic scale) using different approaches for splitting disconnected communities with Parallel Louvain algorithm]{
    \label{fig:optlousp--disconnected}
    \includegraphics[width=0.98\linewidth]{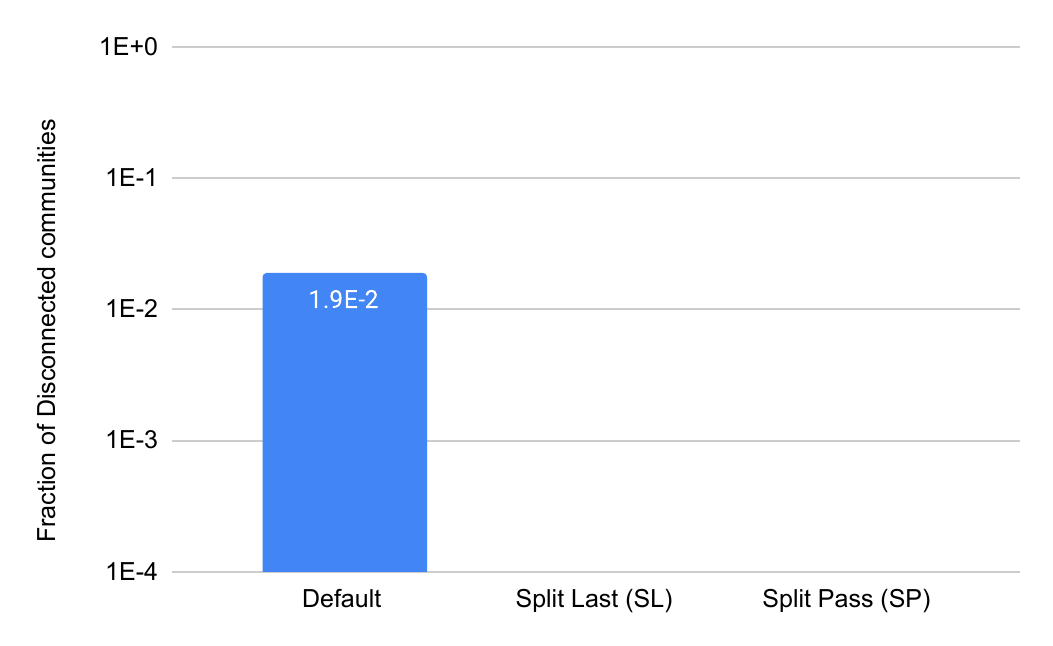}
  } \\[-2ex]
  \caption{Mean relative runtime, modularity, and fraction of disconnected communities (log-scale) using \textit{Split Last (SL)} and \textit{Split Pass (SP)} approaches for splitting disconnected communities with Parallel Louvain algorithm\ignore{\cite{sahu2023gveleiden}} across all graphs in the dataset. Both \textit{SL} and \textit{SP} approaches employ\ignore{minimum label-based} \textit{Label Propagation (LP)}, \textit{Label Propagation with Pruning (LPP)}, or \textit{Breadth First Search (BFS)} techniques for splitting.}
  \label{fig:optlousp}
\end{figure}

%% file: src/alg-louvain.tex
\begin{algorithm}[hbtp]
\caption{GSP-Louvain: Our Parallel Louvain algorithm which identifies communities that are not internally disconnected.}
\label{alg:louvain}
\begin{algorithmic}[1]
\Require{$G(V, E)$: Input graph}
\Ensure{$G'(V', E')$: Input/super-vertex graph}
\Ensure{$C$: Community membership of each vertex}
\Ensure{$C'$: Community membership of each vertex in $G'$}
\Ensure{$K'$: Total edge weight of each vertex}
\Ensure{$\Sigma'$: Total edge weight of each community}
\Ensure{$l_i$: Number of iterations performed (per pass)}
\Ensure{$l_p$: Number of passes performed}
\Ensure{$\tau, \tau_{agg}$: Iteration, Aggregation tolerance}

\Statex

\Function{louvain}{$G$} \label{alg:louvain--begin}
  \State Vertex membership: $C \gets [0 .. |V|)$ \textbf{;} $G' \gets G$ \label{alg:louvain--initialization}
  \ForAll{$l_p \in [0 .. \text{\small{MAX\_PASSES}})$} \label{alg:louvain--passes-begin}
    \State $\Sigma' \gets K' \gets vertexWeights(G')$ \textbf{;} $C' \gets [0 .. |V'|)$ \label{alg:louvain--reset-weights}
    \State Mark all vertices in $G'$ as unprocessed \label{alg:louvain--reset-affected}
    \State $l_i \gets louvainMove(G', C', K', \Sigma', \tau)$ \Comment{Algorithm \ref{alg:commonlm}} \label{alg:louvain--local-move}
    \State $C' \gets splitDisconnectedBfs(G', C')$ \Comment{Algorithm \ref{alg:splitbfs}} \label{alg:louvain--split}
    \If{$l_i \le 1$} \textbf{break} \Comment{Globally converged?} \label{alg:louvain--globally-converged}
    \EndIf
    \State $|\Gamma|, |\Gamma_{old}| \gets$ Number of communities in $C$, $C'$
    \If{$|\Gamma|/|\Gamma_{old}| > \tau_{agg}$} \textbf{break} \Comment{Low shrink?} \label{alg:louvain--aggregation-tolerance}
    \EndIf
    \State $C' \gets$ Renumber communities in $C'$ \label{alg:louvain--renumber}
    \State $C \gets$ Lookup dendrogram using $C$ to $C'$ \label{alg:louvain--lookup}
    \State $G' \gets louvainAggregate(G', C')$ \Comment{Algorithm \ref{alg:commonag}} \label{alg:louvain--aggregate}
    \State $\tau \gets \tau / \text{\small{TOLERANCE\_DROP}}$ \Comment{Threshold scaling} \label{alg:louvain--threshold-scaling}
  \EndFor \label{alg:louvain--passes-end}
  \State $C \gets$ Lookup dendrogram using $C$ to $C'$ \label{alg:louvain--lookup-last}
  \Return{$C$} \label{alg:louvain--return}
\EndFunction \label{alg:louvain--end}
\end{algorithmic}
\end{algorithm}

%% file: 05-evaluation.tex
\subsection{Experimental Setup}
\label{sec:setup}

\subsubsection{System used}

We utilize a server comprising two Intel Xeon Gold 6226R processors, with each processor housing $16$ cores operating at $2.90$ GHz. Each core is equipped with a $1$ MB L1 cache, a $16$ MB L2 cache, and a shared L3 cache of $22$ MB. The system is configured with $376$ GB of RAM and is running CentOS Stream 8.

\subsubsection{Configuration}

We employ 32-bit integers to represent vertex IDs and 32-bit floats for edge weights, while computations and hashtable values utilize 64-bit floats. We utilize $64$ threads to match the number of available cores on the system, unless stated otherwise. Compilation is performed using GCC 8.5 and OpenMP 4.5.

\subsubsection{Dataset}
\label{sec:dataset}

The graphs utilized in our experiments are listed in Table \ref{tab:dataset}, sourced from the SuiteSparse Matrix Collection \cite{suite19}. These graphs exhibit $3.07$ to $214$ million vertices, and $25.4$ million to $3.80$ billion edges. We ensure that the edges are undirected and weighted, with a default weight of $1$.

\input{src/tab-dataset}

\subsection{Performance Comparison}
\label{sec:comparison}

We now compare the performance of GSP-Louvain with the original Leiden \cite{com-traag19}, NetworKit Leiden \cite{staudt2016networkit}, and cuGraph Leiden \cite{kang2023cugraph}. For the original Leiden, we employ a C++ program to initialize a \texttt{ModularityVertexPartition} upon the loaded graph and invoke \texttt{optimise\_partition()} to determine the community membership of each vertex. On graphs with high edge counts, such as \textit{webbase-2001} and \textit{sk-2005}, utilizing \texttt{ModularityVertexPartition} can result in disconnected communities due to numerical precision issues \cite{traag2024leiden}. Despite a positive improvement in separating disconnected parts, the large total edge weight of the graph can render it effectively near zero. For such graphs, we employ \texttt{RBConfigurationVer} \texttt{texPartition}, as it uses unscaled modularity improvements, avoiding the occurrence of disconnected communities. In the case of NetworKit Leiden, we create a Python script to call \texttt{ParallelLeiden()}, while constraining the number of passes to $10$. For each graph, we measure the runtime of each implementation and the modularity of the resulting communities five times to obtain an average. For cuGraph Leiden, we create a Python script that configures the Rapids Memory Manager (RMM) to use a pool allocator, enabling faster memory allocations. We then execute \texttt{cugraph.leiden()} on the loaded graph. For each graph, we record the runtime and measure the modularity of the resulting communities, repeating the process five times to average the results. The runtime of the first run is excluded to ensure subsequent measurements reflect RMM's pool usage without the overhead from initial CUDA memory allocations. Additionally, we store the community membership vector (for each vertex in the graph) in a file and subsequently determine the number of disconnected components using Algorithm \ref{alg:disconnected}. Throughout these evaluations, we optimize for modularity as the quality function.

Figure \ref{fig:cmplousp--runtime} presents the runtimes of the original Leiden, NetworKit Leiden, cuGraph Leiden, and GSP-Louvain on each graph in the dataset. On the \textit{sk-2005} graph, GSP-Louvain identifies communities in $8.9$ seconds, achieving a processing rate of $410$ million edges/s. Figure \ref{fig:cmplousp--speedup} illustrates the speedup of GSP-Louvain relative to the original Leiden, NetworKit Leiden, and cuGraph Leiden. On average, GSP-Louvain exhibit speedups of $391\times$, $6.9\times$, and $2.6\times$, respectively. Figure \ref{fig:cmplousp--modularity} displays the modularity of communities obtained using each implementation. On average, GSP-Louvain achieves modularity values that are $0.3\%$ lower than those obtained by the original Leiden and cuGraph Leiden, but $25\%$ higher than those obtained by NetworKit Leiden (particularly noticeable on road networks and protein k-mer graphs), and $3.5\%$ higher modularity that cuGraph Leiden (mainly because cuGraph Leiden fails to run on web graphs, which are well-clusterable). Finally, Figure \ref{fig:cmplousp--disconnected} illustrates the fraction of disconnected communities obtained by each implementation. The absence of bars indicates the absence of disconnected communities. Communities identified by the original Leiden and GSP-Louvain exhibit no disconnected communities. However, on average, NetworKit Leiden and cuGraph Leiden exhibit fractions of disconnected communities amounting to $1.5\times10^{-2}$ and $6.6\times10^{-5}$, respectively, particularly noticeable on web graphs and social networks. This is likely due to error in their implementation.\ignore{Thus, GSP-Louvain effectively tackles the issue of disconnected communities, while being significantly faster than existing alternatives, and attaining similar modularity scores.} Figure \ref{fig:cmpduo} depicts the comparison of GVE-Louvain and GSP-Louvain. This comparison is explained in detail in Section \ref{sec:comparison-extra}.

\input{src/fig-cmplousp}
\input{src/fig-splitlousp}
\input{src/fig-scaling}

\subsection{Performance Analysis}
\label{sec:analysis}

We proceed to analyze the performance of GSP-Louvain. The phase-wise and pass-wise split of GSP-Louvain is depicted in Figures \ref{fig:splitlousp--phase} and \ref{fig:splitlousp--pass}. Figure \ref{fig:splitlousp--phase} illustrates that GSP-Louvain devotes a considerable portion of its runtime to the local-moving phase on web graphs, road networks, and protein k-mer graphs, while it predominantly focuses on the aggregation phase on social networks, and on the splitting phase on road networks. The pass-wise breakdown for GSP-Louvain, shown in Figure \ref{fig:splitlousp--pass}, indicates that the initial pass is computationally intensive for high-degree graphs (such as web graphs and social networks), while subsequent passes take precedence in terms of execution time on low-degree graphs (such as road networks and protein k-mer graphs).

On average, GSP-Louvain dedicates $40\%$ of its runtime to the local-moving phase, $25\%$ to the splitting phase, $24\%$ to the aggregation phase, and $11\%$ to other steps (including initialization, renumbering communities, dendrogram lookup, and resetting communities). Additionally, the first pass of GSP-Louvain accounts for $73\%$ of the total runtime. This initial pass in GSP-Louvain is computationally demanding due to the size of the original graph (subsequent passes operate on super-vertex graphs).

\subsection{Strong Scaling}
\label{sec:scaling}

Finally, we evaluate the strong scaling performance of GSP-Louvain by varying the number of threads from $1$ to $64$ (in multiples of $2$) for each input graph. We measure the total time taken for GSP-Louvain to identify communities, with its respective phase splits, including local-moving, splitting, aggregation, and other associated steps. The strong scaling of GSP-Louvain is illustrated in Figure \ref{fig:scaling}.

With $32$ threads, GSP-Louvain achieve an average speedup of $8.4\times$, compared to single-threaded execution. This indicates a performance increase of $1.5\times$ for every doubling of threads. The scalability is limited due to the sequential nature of steps/phases in the algorithm, as well as the lower scalability of splitting and aggregation phases. At $64$ threads, GSP-Louvain is impacted by NUMA effects, resulting in speedups of only $9.4\times$.

%% file: src/tab-dataset.tex
\begin{table}[hbtp]
  \centering
  \caption{List of $13$ graphs obtained SuiteSparse Matrix Collection \cite{suite19} (directed graphs are marked with $*$). Here, $|V|$ is the number of vertices, $|E|$ is the number of edges (after adding reverse edges), $D_{avg}$ is the average degree, and $|\Gamma|$ is the number of communities obtained with \textit{GSP-Louvain}.\ignore{In the table, B refers to a billion, M refers to a million and K refers a thousand.}}
  \label{tab:dataset}
  \begin{tabular}{|c||c|c|c|c|}
    \toprule
    \textbf{Graph} &
    \textbf{\textbf{$|V|$}} &
    \textbf{\textbf{$|E|$}} &
    \textbf{\textbf{$D_{avg}$}} &
    \textbf{\textbf{$|\Gamma|$}} \\
    % \textbf{$1 - \Gamma_G$} \\
    \midrule
    \multicolumn{5}{|c|}{\textbf{Web Graphs (LAW)}} \\ \hline
    indochina-2004$^*$ & 7.41M & 341M & 41.0 & 4.28K \\ \hline  % & \num{4.7e-4} & 2.9 GB
    uk-2002$^*$ & 18.5M & 567M & 16.1 & 42.8K \\ \hline  % & \num{9.6e-5} & 16 GB
    arabic-2005$^*$ & 22.7M & 1.21B & 28.2 & 3.58K \\ \hline  % & \num{5.5e-4} & 11 GB
    uk-2005$^*$ & 39.5M & 1.73B & 23.7 & 20.4K \\ \hline  % & \num{9.6e-5} & 16 GB
    webbase-2001$^*$ & 118M & 1.89B & 8.6 & 2.77M \\ \hline  % & \num{7.3e-7} & 18 GB
    it-2004$^*$ & 41.3M & 2.19B & 27.9 & 5.10K \\ \hline  % & \num{3.8e-4} & 19 GB
    sk-2005$^*$ & 50.6M & 3.80B & 38.5 & 3.86K \\ \hline  % & \num{5.8e-4} & 33 GB
    \multicolumn{5}{|c|}{\textbf{Social Networks (SNAP)}} \\ \hline
    com-LiveJournal & 4.00M & 69.4M & 17.4 & 4.47K \\ \hline  % & \num{7.9e-4} & 480 MB
    com-Orkut & 3.07M & 234M & 76.2 & 43 \\ \hline  % & \num{6.7e-2} & 1.7 GB
    \multicolumn{5}{|c|}{\textbf{Road Networks (DIMACS10)}} \\ \hline
    asia\_osm & 12.0M & 25.4M & 2.1 & 2.50K \\ \hline  % & \num{8.4e-4} & 200 MB
    europe\_osm & 50.9M & 108M & 2.1 & 3.36K \\ \hline  % & \num{6.6e-4} & 910 MB
    \multicolumn{5}{|c|}{\textbf{Protein k-mer Graphs (GenBank)}} \\ \hline
    kmer\_A2a & 171M & 361M & 2.1 & 19.8K \\ \hline  % & \num{9.4e-5} & 3.2 GB
    kmer\_V1r & 214M & 465M & 2.2 & 7.37K \\ \hline  % & \num{3.2e-4} & 4.2 GB
  \bottomrule
  \end{tabular}
\end{table}

%% file: src/fig-cmplousp.tex
\begin{figure*}[hbtp]
  \centering
  \subfigure[Runtime in seconds (logarithmic scale) with \textit{Original Leiden}, \textit{NetworKit Leiden}, \textit{cuGraph Leiden}, and \textit{GSP-Louvain}]{
    \label{fig:cmplousp--runtime}
    \includegraphics[width=0.98\linewidth]{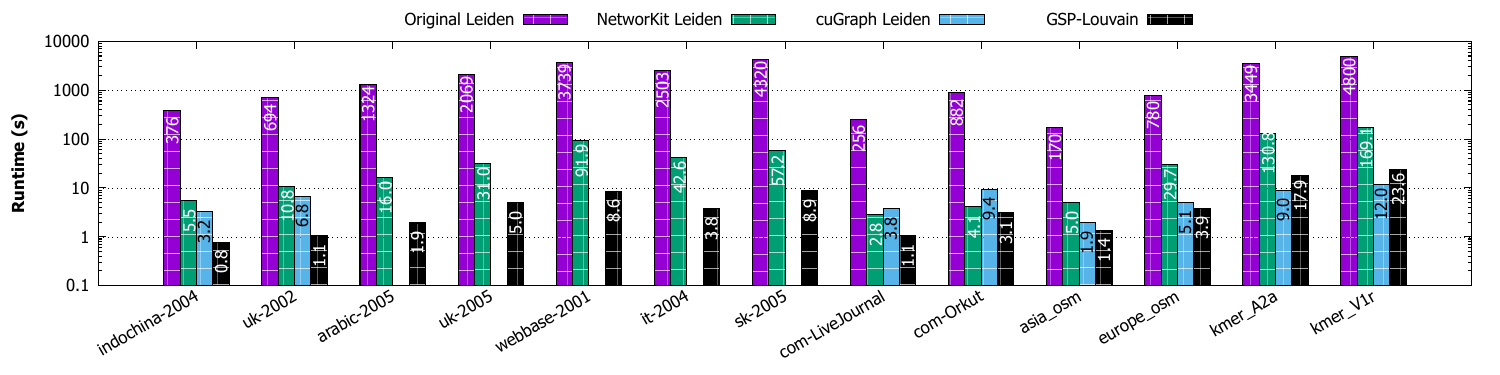}
  } \\[-0ex]
  \subfigure[Speedup of \textit{GSP-Louvain} (logarithmic scale) with respect to \textit{Original Leiden}, \textit{NetworKit Leiden}, and \textit{cuGraph Leiden}]{
    \label{fig:cmplousp--speedup}
    \includegraphics[width=0.98\linewidth]{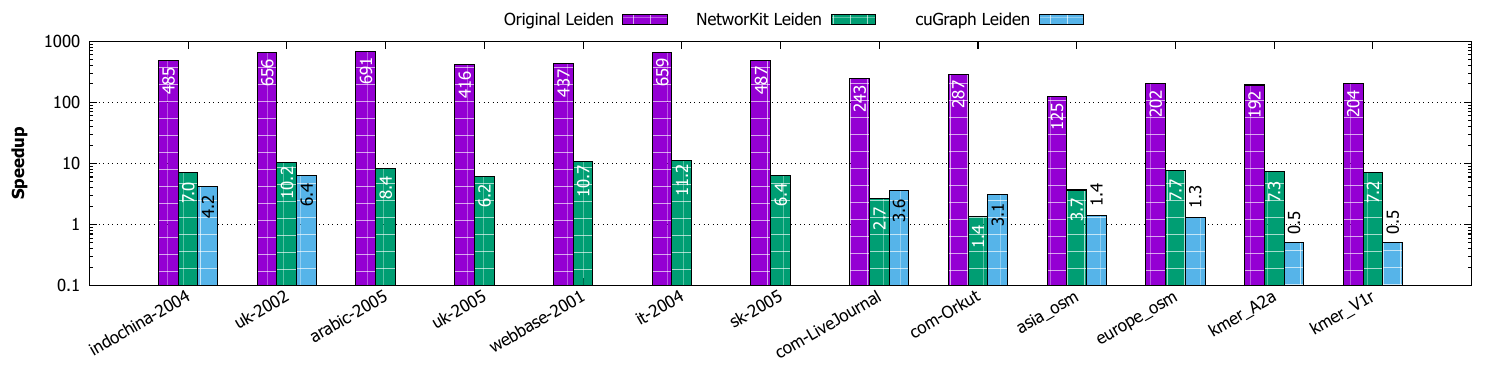}
  } \\[-0ex]
  \subfigure[Modularity of communities obtained with \textit{Original Leiden}, \textit{NetworKit Leiden}, \textit{cuGraph Leiden}, and \textit{GSP-Louvain}]{
    \label{fig:cmplousp--modularity}
    \includegraphics[width=0.98\linewidth]{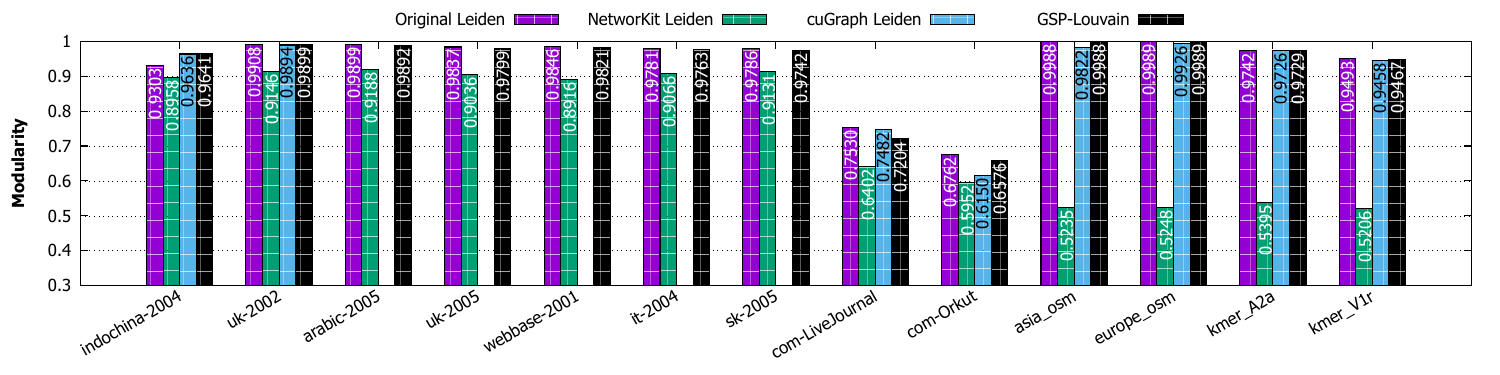}
  } \\[-0ex]
  \subfigure[Fraction of disconnected communities (logarithmic scale) with \textit{Original Leiden}, \textit{NetworKit Leiden}, \textit{cuGraph Leiden}, and \textit{GSP-Louvain}]{
    \label{fig:cmplousp--disconnected}
    \includegraphics[width=0.98\linewidth]{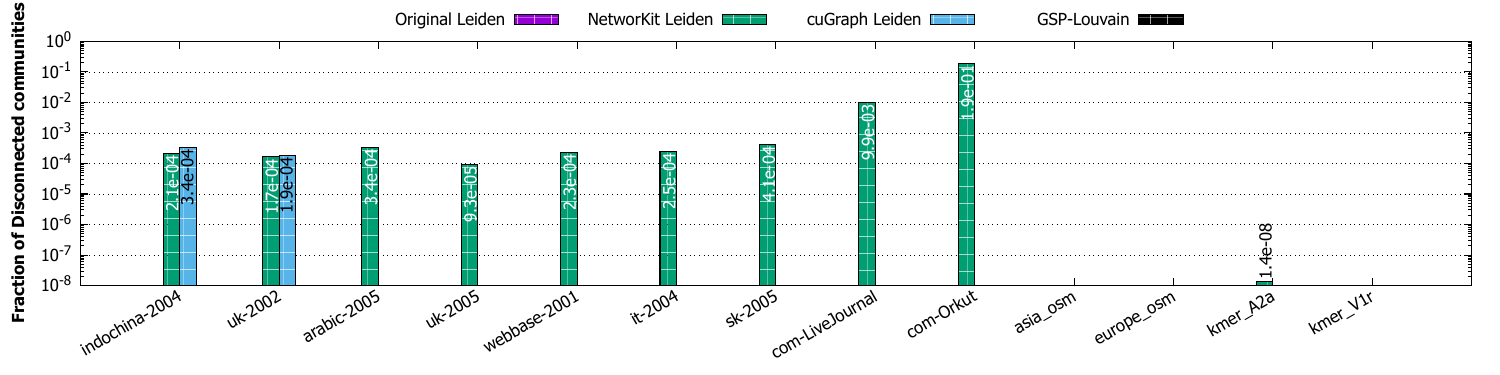}
  } \\[-2ex]
  \caption{Runtime in seconds (log-scale), speedup (log-scale), modularity, and fraction of disconnected communities (log-scale) with \textit{Original Leiden}, \textit{NetworKit Leiden}, \textit{cuGraph Leiden}, and \textit{GSP-Louvain} for each graph in the dataset.}
  \label{fig:cmplousp}
\end{figure*}

%% file: src/fig-splitlousp.tex
\begin{figure*}[hbtp]
  \centering
  \subfigure[Phase split of \textit{GSP-Louvain}]{
    \label{fig:splitlousp--phase}
    \includegraphics[width=0.48\linewidth]{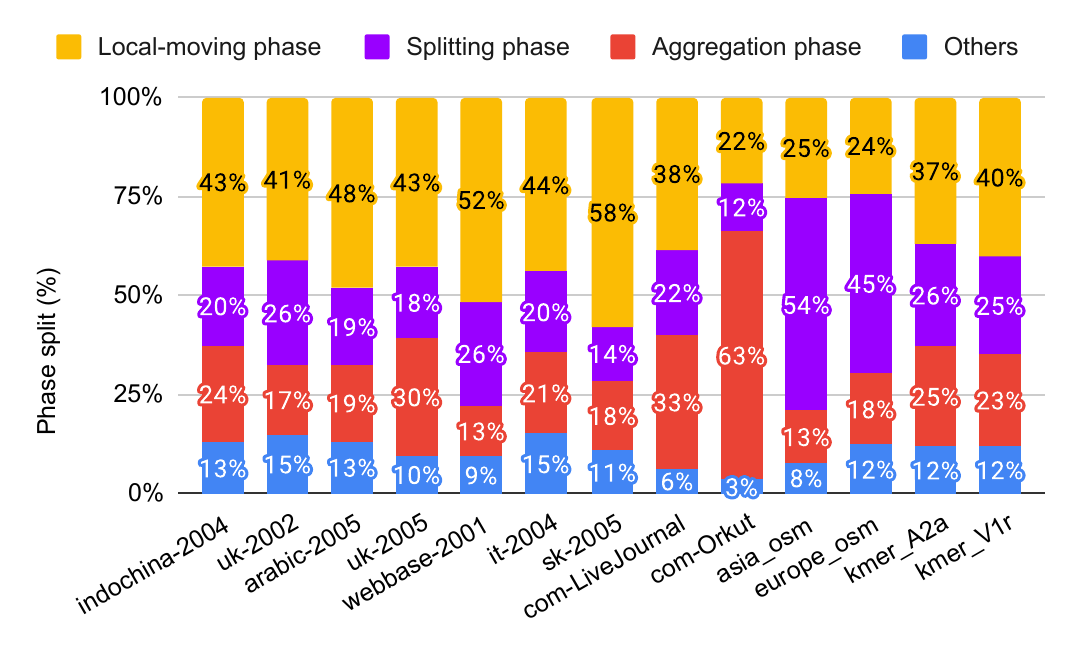}
  }
  \subfigure[Pass split of \textit{GSP-Louvain}]{
    \label{fig:splitlousp--pass}
    \includegraphics[width=0.48\linewidth]{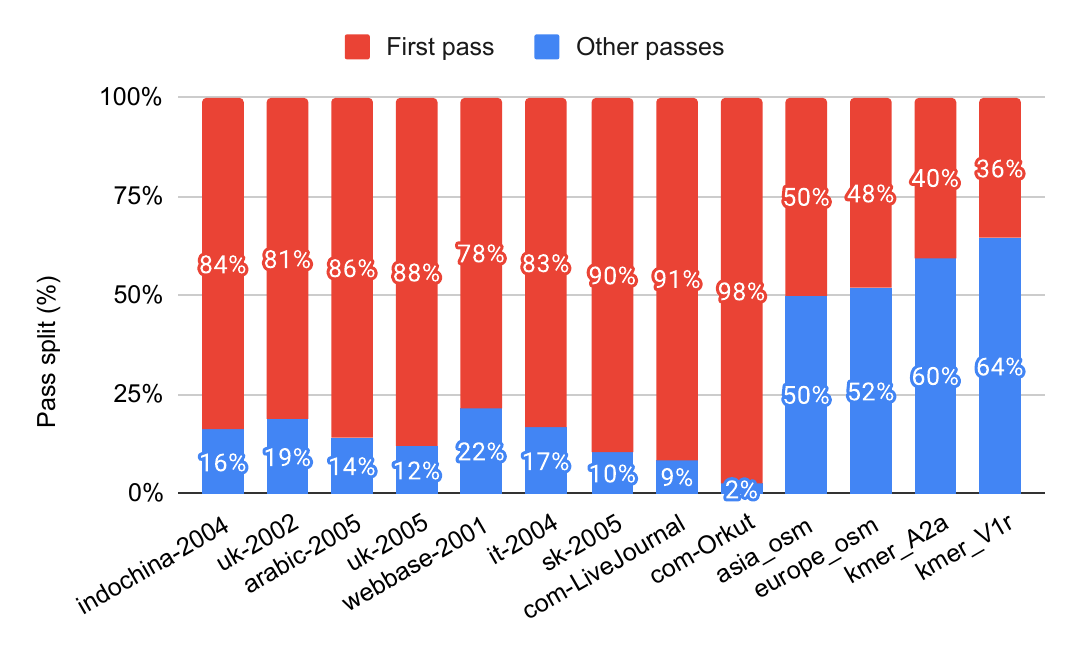}
  } \\[-2ex]
  \caption{Phase split of \textit{GSP-Louvain} shown on the left, and pass split shown on the right for each graph in the dataset.}
  \label{fig:splitlousp}
\end{figure*}

%% file: src/fig-scaling.tex
\begin{figure}[hbtp]
  \centering
  \subfigure{
    \label{fig:scaling--lousp}
    \includegraphics[width=0.98\linewidth]{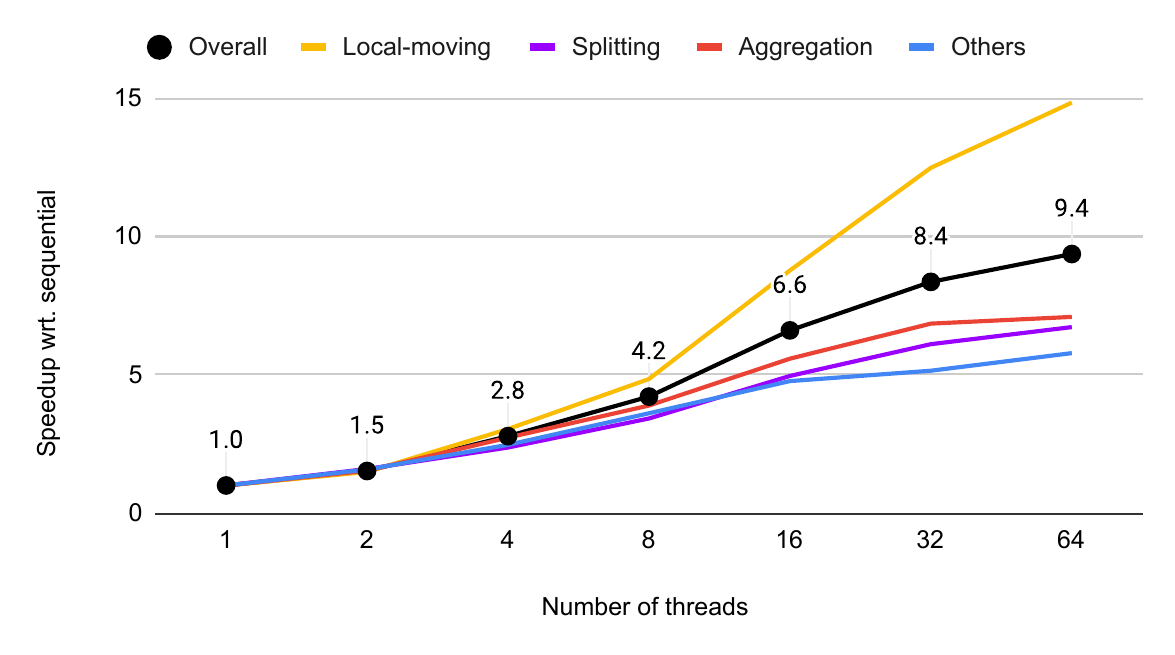}
  } \\[-2ex]
  \caption{Overall speedup of \textit{GSP-Louvain}, and its various phases (local-moving, splitting, aggregation, and others), with increasing number of threads (in multiples of 2).}
  \label{fig:scaling}
\end{figure}

%% file: 06-conclusion.tex
In this study, we proposed GSP-Louvain, another approach to mitigate the issue of disconnected communities with the Louvain algorithm. Utilizing a system featuring two 16-core Intel Xeon Gold 6226R processors, we demonstrated that GSP-Louvain not only rectifies this issue but also achieves an impressive processing rate of $410 M$ edges/s on a $3.8 B$ edge graph. Comparatively, its surpasses the original Leiden, NetworKit Leiden, and cuGraph Leiden by $391\times$, $6.9\times$, and $2.6\times$, respectively. Moreover, the identified communities exhibit similar quality to the original Leiden, and are $25\%$ and $3.5\%$ higher in quality than those produced by NetworKit and cuGraph, respectively. Additionally, GSP-Louvain exhibits a performance improvement rate of $1.5\times$ for every doubling of threads.

In this version of this report, we addressed issues in measuring disconnected communities for the original Leiden and cuGraph Leiden, which arose due to the number of vertices in a graph varying between the Matrix Market and the Edgelist formats (which does not have isolated vertices), and used the \texttt{RBConfigurationVertexPart} \texttt{ition} with the original Leiden for large graphs (i.e., \textit{webbase-2001} and \textit{sk-2005}). Further, we removed any discussion on GSP-Leiden (it was similar to GSP-Louvain, but adapted to the Leiden algorithm).

%% file: aa-appendix.tex
\subsection{Phases of GSP-Louvain}

Here, we explain the local-moving, and aggregation phases of GSP-Louvain. Details on the splitting phase are in Section \ref{sec:splitbfs}.

\subsubsection{Local-moving phase of GSP-Louvain}
\label{sec:commonlm}

The pseudocode for the local-moving phase of GSP-Louvain is outlined in Algorithm \ref{alg:commonlm}. Here, vertices are iteratively moved between communities to maximize modularity. The \texttt{louvainMove()} function takes as input the current graph $G'$, community membership $C'$, total edge weight of each vertex $K'$ and each community $\Sigma'$, and the iteration tolerance $\tau$. It returns the number of iterations performed $l_i$.

Lines \ref{alg:commonlm--iterations-begin}-\ref{alg:commonlm--iterations-end} encapsulate the primary loop of the local-moving phase. Initially, all vertices are designated as unprocessed (line \ref{alg:commonlm--reset-affected}). Subsequently, in line \ref{alg:commonlm--init-deltaq}, the total delta-modularity per iteration $\Delta Q$ is initialized. Next, parallel iteration over unprocessed vertices is conducted (lines \ref{alg:commonlm--loop-vertices-begin}-\ref{alg:commonlm--loop-vertices-end}). For each unprocessed vertex $i$, $i$ is flagged as processed, i.e., vertex pruning (line \ref{alg:commonlm--prune}), followed by scanning communities connected to $i$, excluding itself (line \ref{alg:commonlm--scan}). Further, the best community $c^*$ for moving $i$ to is determined (line \ref{alg:commonlm--best-community-begin}), and the delta-modularity $\delta Q^*$ of moving $i$ to $c^*$ is computed (line \ref{alg:commonlm--best-community-end}). If a superior community is identified (with $\delta Q^* > 0$), the community membership of $i$ is updated (lines \ref{alg:commonlm--perform-move-begin}-\ref{alg:commonlm--perform-move-end}), and its neighbors are marked as unprocessed (line \ref{alg:commonlm--remark}). If not, $i$ stays in its original community. Line \ref{alg:commonlm--locally-converged} verifies if the local-moving phase has converged, terminating the loop if so (or if $MAX\_ITERATIONS$ is reached). Finally, in line \ref{alg:commonlm--return}, the number of iterations performed $l_i$ is returned.

\input{src/alg-commonlm}

\subsubsection{Aggregation phase of GSP-Louvain}
\label{sec:commonag}

Finally, we provide the pseudocode for the aggregation phase in Algorithm \ref{alg:commonag}, which aggregates communities into super-vertices in preparation for the subsequent pass of the Louvain algorithm, operating on the super-vertex graph. The \texttt{louvainAggregate()} function accepts the current graph $G'$ and the community membership $C'$ as input and returns the super-vertex graph $G''$.

In lines \ref{alg:commonag--coff-begin}-\ref{alg:commonag--coff-end}, the offsets array for the community vertices Compressed Sparse Row (CSR) $G'{C'}.offsets$ is computed. Initially, this involves determining the number of vertices in each community using \texttt{countCommunityVertices()} and subsequently performing an exclusive scan on the array. Then, in lines \ref{alg:commonag--comv-begin}-\ref{alg:commonag--comv-end}, a parallel iteration over all vertices is conducted to atomically populate vertices belonging to each community into the community graph CSR $G'{C'}$. Following this, the offsets array for the super-vertex graph CSR is determined by estimating the degree of each super-vertex. This process includes calculating the total degree of each community with \texttt{communityTotalDegree()} and performing an exclusive scan on the array (lines \ref{alg:commonag--yoff-begin}-\ref{alg:commonag--yoff-end}). As a result, the super-vertex graph CSR exhibits sparsity, with gaps between the edges and weights arrays of each super-vertex in the CSR. Following that, in lines \ref{alg:commonag--y-begin}-\ref{alg:commonag--y-end}, a parallel iteration over all communities $c \in [0, |\Gamma|)$ is executed. For each vertex $i$ belonging to community $c$, all communities $d$ (along with associated edge weight $w$) linked to $i$, as defined by \texttt{scanCommunities()} in Algorithm \ref{alg:commonlm}, are included in the per-thread hashtable $H_t$. Once $H_t$ is populated with all communities (and their associated weights) linked to community $c$, these are atomically added as edges to super-vertex $c$ in the super-vertex graph $G''$. Finally, in line \ref{alg:commonag--return}, the super-vertex graph $G''$ is returned.

\input{src/alg-commonag}

\subsection{Finding disconnected communities}

We introduce our parallel algorithm designed to identify disconnected communities, given the original graph and the community membership of each vertex. The core principle involves assessing the size of each community, selecting a representative vertex from each community, navigating within the community from that vertex while avoiding neighboring communities, and designating a community as disconnected if all its vertices are unreachable. We investigate four distinct approaches, distinguished by their utilization of parallel Depth-First Search (DFS) or Breadth-First Search (BFS), and whether per-thread or shared \textit{visited} flags are employed. When shared visited flags are utilized, each thread scans all vertices but exclusively processes its designated community based on the community ID. Our findings reveal that employing parallel BFS traversal with a shared flag vector yields the most efficient results. Given the deterministic nature of this algorithm, all approaches yield identical outcomes. Algorithm \ref{alg:disconnected} outlines the pseudocode for this approach. Here, the \texttt{disconnectedCommunities()} function takes the input graph $G$ and the community membership $C$ as input and returns the disconnected flag $D$ for each community.

\input{src/alg-disconnected}

Let us now delve into Algorithm \ref{alg:disconnected}. Initially, in line \ref{alg:disconnected--init}, we initialize the disconnected community flag $D$ and the visited vertices flags $vis$. Line \ref{alg:disconnected--sizes} computes the size of each community $S$ in parallel using the \texttt{communitySizes()} function. Subsequently, each thread processes each vertex $i$ in the graph $G$ in parallel (lines \ref{alg:disconnected--loop-begin}-\ref{alg:disconnected--loop-end}). In line \ref{alg:disconnected--unreached}, we determine the community membership of $i$ ($c$), and set the count of vertices reached from $i$ to $0$. If community $c$ is either empty or not in the work-list of the current thread $work_t$, the thread proceeds to the next iteration (line \ref{alg:disconnected--work}). However, if community $c$ is non-empty and in the work-list of the current thread $work_t$, we perform BFS from vertex $i$ to explore vertices in the same community. This utilizes lambda functions $f_{if}$ to conditionally execute BFS to vertex $j$ if it belongs to the same community, and $f_{do}$ to update the count of reached vertices after each vertex is visited during BFS (line \ref{alg:disconnected--bfs}). If the number of vertices $reached$ during BFS is less than the community size $S[c]$, we mark community $c$ as disconnected (line \ref{alg:disconnected--mark}). Finally, we update the size of the community $S[c]$ to $0$, indicating that the community has been processed (line \ref{alg:disconnected--processed}). It's important to note that the work-list $work_t$ for each thread with ID $t$ is defined as a set containing communities $[t\chi,\ t(\chi+1))\ \cup\ [T\chi + t\chi,\ T\chi + t(\chi+1))\ \cup\ \ldots$, where $\chi$ is the chunk size, and $T$ is the number of threads. In our implementation, we utilize a chunk size of $\chi = 1024$.

\subsection{Additional Performance comparison}
\label{sec:comparison-extra}

We now compare the performance of GSP-Louvain with GVE-Louvain \cite{sahu2023gvelouvain}. Similar to our previous approach, we execute each algorithm five times for each graph\ignore{in the dataset} to mitigate measurement noise, and report the averages in Figures \ref{fig:cmpduo--runtime}, \ref{fig:cmpduo--speedup}, \ref{fig:cmpduo--modularity}, and \ref{fig:cmpduo--disconnected}. 

Figure \ref{fig:cmpduo--runtime} illustrates the runtimes of GSP-Louvain and GVE-Louvain on each graph in the dataset. On average, GSP-Louvain exhibits about a $32\%$ increase in runtime compared to GVE-Louvain. This additional computational time is a compromise made to ensure the absence of internally disconnected communities. Figure \ref{fig:cmpduo--modularity} presents the modularity of communities obtained by each implementation. On average, the modularity of communities obtained using GSP-Louvain and GVE-Louvain remains roughly identical. Lastly, Figure \ref{fig:cmpduo--disconnected} displays the fraction of internally disconnected communities identified by each implementation. Communities obtained with GSP-Louvain exhibit no disconnected communities, whereas communities identified with GVE-Louvain feature on average $3.9\%$ disconnected communities.

\input{src/fig-cmpduo}

%% file: src/alg-commonlm.tex
\begin{algorithm}[hbtp]
\caption{Local-moving phase of GSP-Louvain \cite{sahu2023gvelouvain}.}
\label{alg:commonlm}
\begin{algorithmic}[1]
\Require{$G'(V', E')$: Input/super-vertex graph}
\Require{$C'$: Community membership of each vertex}
\Require{$K'$: Total edge weight of each vertex}
\Require{$\Sigma'$: Total edge weight of each community}
\Require{$\tau$: Iteration tolerance}
\Ensure{$H_t$: Collision-free per-thread hashtable}
\Ensure{$l_i$: Number of iterations performed}

\Statex

\Function{louvainMove}{$G', C', K', \Sigma', \tau$} \label{alg:commonlm--move-begin}
  \State Mark all vertices in $G'$ as unprocessed \label{alg:commonlm--reset-affected}
  \ForAll{$l_i \in [0 .. \text{\small{MAX\_ITERATIONS}})$} \label{alg:commonlm--iterations-begin}
    \State Total delta-modularity per iteration: $\Delta Q \gets 0$ \label{alg:commonlm--init-deltaq}
    \ForAll{unprocessed $i \in V'$ \textbf{in parallel}} \label{alg:commonlm--loop-vertices-begin}
      \State Mark $i$ as processed (prune) \label{alg:commonlm--prune}
      \State $H_t \gets scanCommunities(\{\}, G', C', i, false)$ \label{alg:commonlm--scan}
      \State $\rhd$ Use $H_t, K', \Sigma'$ to choose best community
      \State $c^* \gets$ Best community linked to $i$ in $G'$ \label{alg:commonlm--best-community-begin}
      \State $\delta Q^* \gets$ Delta-modularity of moving $i$ to $c^*$ \label{alg:commonlm--best-community-end}
      \If{$c^* = C'[i]$} \textbf{continue} \label{alg:commonlm--best-community-same}
      \EndIf
      \State $\Sigma'[C'[i]] -= K'[i]$ \textbf{;} $\Sigma'[c^*] += K'[i]$ \textbf{atomic} \label{alg:commonlm--perform-move-begin}
      \State $C'[i] \gets c^*$ \textbf{;} $\Delta Q \gets \Delta Q + \delta Q^*$ \label{alg:commonlm--perform-move-end}
      \State Mark neighbors of $i$ as unprocessed \label{alg:commonlm--remark}
    \EndFor \label{alg:commonlm--loop-vertices-end}
    \If{$\Delta Q \le \tau$} \textbf{break} \Comment{Locally converged?} \label{alg:commonlm--locally-converged}
    \EndIf
  \EndFor \label{alg:commonlm--iterations-end}
  \Return{$l_i$} \label{alg:commonlm--return}
\EndFunction \label{alg:commonlm--move-end}

\Statex

\Function{scanCommunities}{$H_t, G', C', i, self$}
  \ForAll{$(j, w) \in G'.edges(i)$}
    \If{\textbf{not} $self$ \textbf{and} $i = j$} \textbf{continue}
    \EndIf
    \State $H_t[C'[j]] \gets H_t[C'[j]] + w$
  \EndFor
  \Return{$H_t$}
\EndFunction
\end{algorithmic}
\end{algorithm}

%% file: src/alg-commonag.tex
\begin{algorithm}[hbtp]
\caption{Aggregation phase of GSP-Louvain \cite{sahu2023gvelouvain}.}
\label{alg:commonag}
\begin{algorithmic}[1]
\Require{$G'(V', E')$: Input/super-vertex graph}
\Require{$C'$: Community membership of each vertex}
\Ensure{$G'_{C'}(V'_{C'}, E'_{C'})$: Community vertices (CSR)}
\Ensure{$G''(V'', E'')$: Super-vertex graph (weighted CSR)}
\Ensure{$*.offsets$: Offsets array of a CSR graph}
\Ensure{$H_t$: Collision-free per-thread hashtable}

\Statex

\Function{louvainAggregate}{$G', C'$}
  \State $\rhd$ Obtain vertices belonging to each community
  \State $G'_{C'}.offsets \gets countCommunityVertices(G', C')$ \label{alg:commonag--coff-begin}
  \State $G'_{C'}.offsets \gets exclusiveScan(G'_{C'}.offsets)$ \label{alg:commonag--coff-end}
  \ForAll{$i \in V'$ \textbf{in parallel}} \label{alg:commonag--comv-begin}
    \State Add edge $(C'[i], i)$ to CSR $G'_{C'}$ atomically
  \EndFor \label{alg:commonag--comv-end}
  \State $\rhd$ Obtain super-vertex graph
  \State $G''.offsets \gets communityTotalDegree(G', C')$ \label{alg:commonag--yoff-begin}
  \State $G''.offsets \gets exclusiveScan(G''.offsets)$ \label{alg:commonag--yoff-end}
  \State $|\Gamma| \gets$ Number of communities in $C'$
  \ForAll{$c \in [0, |\Gamma|)$ \textbf{in parallel}} \label{alg:commonag--y-begin}
    \If{degree of $c$ in $G'_{C'} = 0$} \textbf{continue}
    \EndIf
    \State $H_t \gets \{\}$
    \ForAll{$i \in G'_{C'}.edges(c)$}
      \State $H_t \gets scanCommunities(H, G', C', i, true)$
    \EndFor
    \ForAll{$(d, w) \in H_t$}
      \State Add edge $(c, d, w)$ to CSR $G''$ atomically
    \EndFor
  \EndFor \label{alg:commonag--y-end}
  \Return $G''$ \label{alg:commonag--return}
\EndFunction
\end{algorithmic}
\end{algorithm}

%% file: src/alg-disconnected.tex
\begin{algorithm}[hbtp]
\caption{Finding disconnected communities in parallel \cite{sahu2023gveleiden}.}
\label{alg:disconnected}
\begin{algorithmic}[1]
\Require{$G(V, E)$: Input graph}
\Require{$C$: Community membership of each vertex}
\Ensure{$D$: Disconnected flag for each community}
\Ensure{$S$: Size of each community}
\Ensure{$f_{if}$: Perform BFS to vertex $j$ if condition satisfied}
\Ensure{$f_{do}$: Perform operation after each vertex is visited}
\Ensure{$reached$: Number of vertices reachable from $i$ to $i$'s community}
\Ensure{$vis$: Visited flag for each vertex}
\Ensure{$work_t$: Work-list of current thread}

\Statex

\Function{disconnectedCommunities}{$G, C$} \label{alg:disconnected--begin}
  \State $D \gets \{\}$ \textbf{;} $vis \gets \{\}$ \label{alg:disconnected--init}
  \State $S \gets communitySizes(G, C)$ \label{alg:disconnected--sizes}
  \ForAll{\textbf{threads in parallel}} \label{alg:disconnected--threads-begin}
    \ForAll{$i \in V$} \label{alg:disconnected--loop-begin}
      \State $c \gets C[i]$ \textbf{;} $reached \gets 0$ \label{alg:disconnected--unreached}
      \State $\rhd$ Skip if community $c$ is empty, or
      \State $\rhd$ does not belong to work-list of current thread.
      \If{$S[c] = 0$ \textbf{or} $c \notin work_t$} \textbf{continue} \label{alg:disconnected--work}
      \EndIf
      \State $f_{if} \gets (j) \implies C[j] = c$
      \State $f_{do} \gets (j) \implies reached \gets reached + 1$
      \State $bfsVisitForEach(vis, G, i, f_{if}, f_{do})$ \label{alg:disconnected--bfs}
      \If{$reached < S[c]$} $D[c] \gets 1$ \label{alg:disconnected--mark}
      \EndIf
      \State $S[c] \gets 0$ \label{alg:disconnected--processed}
    \EndFor \label{alg:disconnected--loop-end}
  \EndFor \label{alg:disconnected--threads-end}
  \Return{$D$}
\EndFunction \label{alg:disconnected--end}
\end{algorithmic}
\end{algorithm}

%% file: src/fig-cmpduo.tex
\begin{figure*}[hbtp]
  \centering
  \subfigure[Runtime in seconds (logarithmic scale) with \textit{GVE-Louvain} and \textit{GSP-Louvain}]{
    \label{fig:cmpduo--runtime}
    \includegraphics[width=0.98\linewidth]{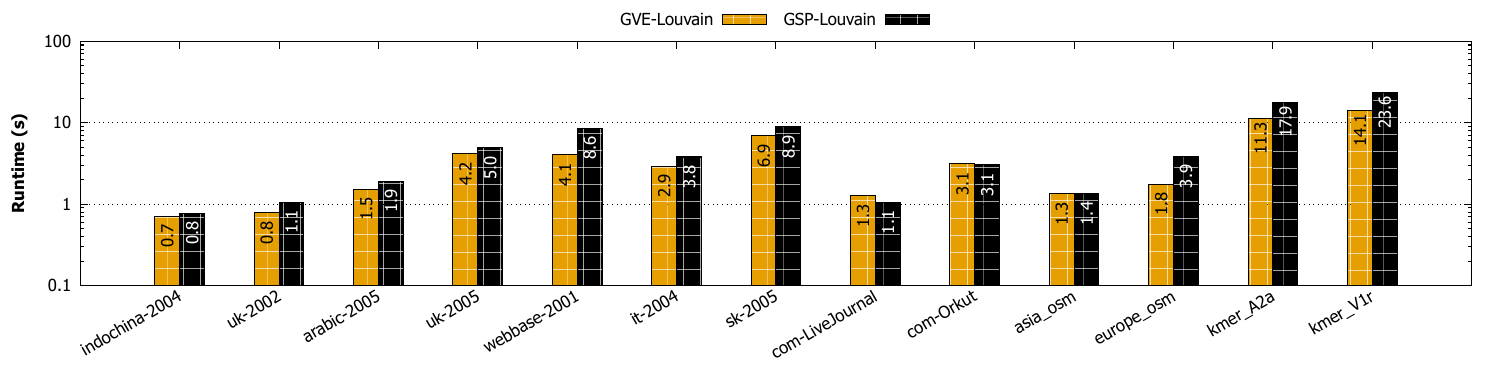}
  } \\[-0ex]
  \subfigure[Speedup of \textit{GSP-Louvain} with respect to \textit{GVE-Louvain}.]{
    \label{fig:cmpduo--speedup}
    \includegraphics[width=0.98\linewidth]{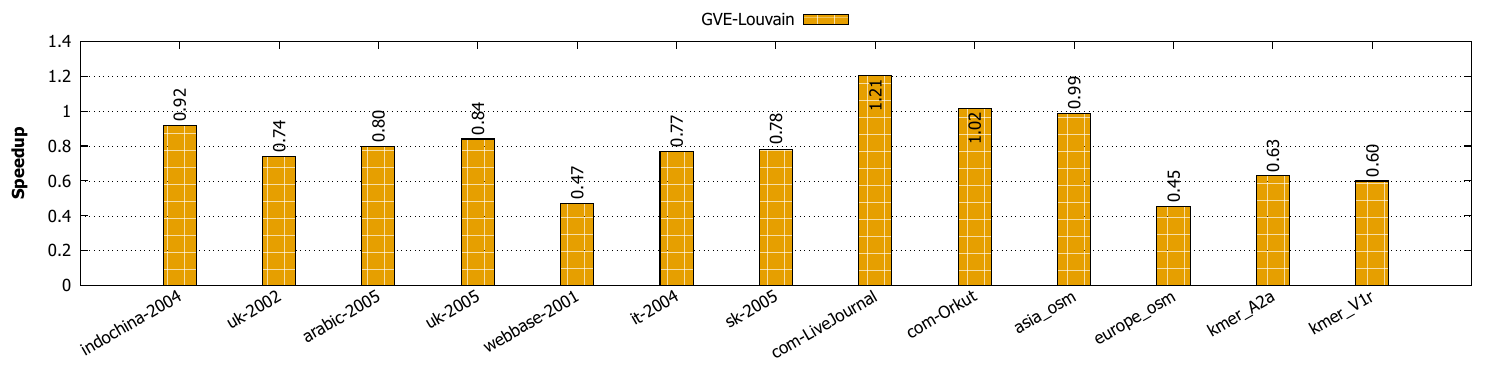}
  } \\[-0ex]
  \subfigure[Modularity of communities obtained with \textit{GVE-Louvain} and \textit{GSP-Louvain}.]{
    \label{fig:cmpduo--modularity}
    \includegraphics[width=0.98\linewidth]{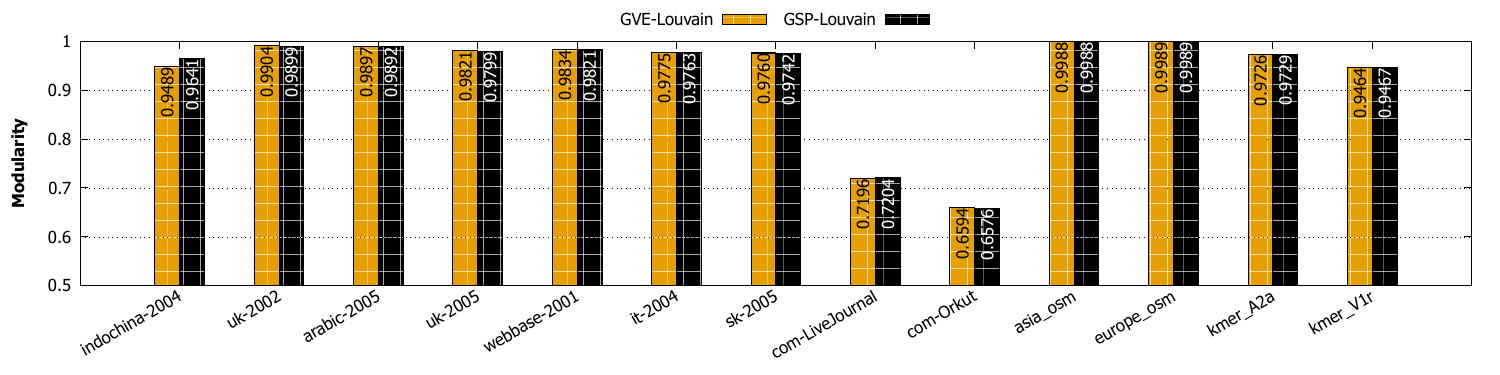}
  } \\[-0ex]
  \subfigure[Fraction of disconnected communities (logarithmic scale) with \textit{GVE-Louvain} and \textit{GSP-Louvain}.]{
    \label{fig:cmpduo--disconnected}
    \includegraphics[width=0.98\linewidth]{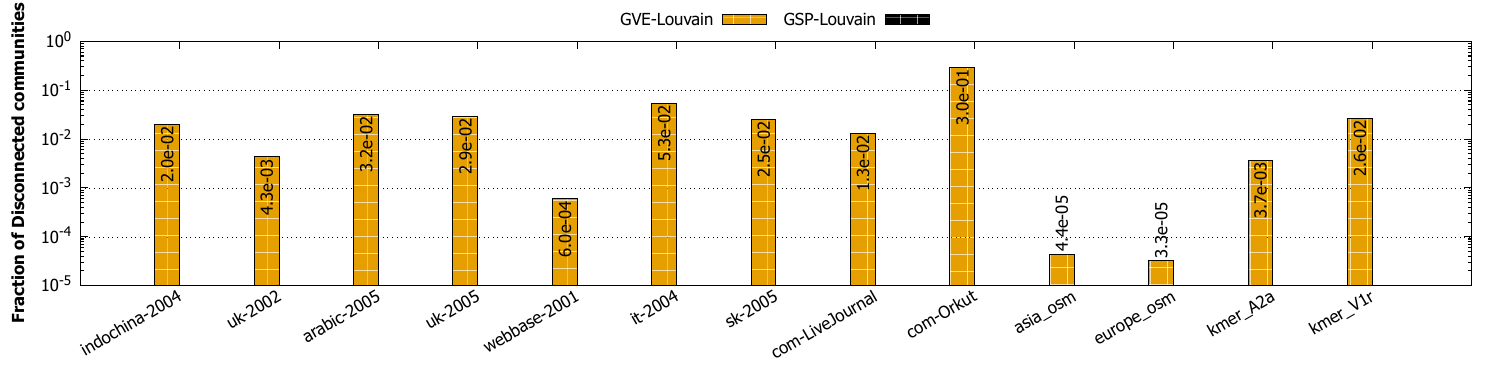}
  } \\[-2ex]
  \caption{Runtime in seconds (log-scale), speedup, modularity, and fraction of disconnected communities (log-scale) with \textit{GVE-Louvain} and \textit{GSP-Louvain} for each graph in the dataset.}
  \label{fig:cmpduo}
\end{figure*}

%% file: main.bbl

\begin{thebibliography}{47}

%%% the Web address.

\ifx \showCODEN    \undefined \def \showCODEN     #1{\unskip}     \fi
\ifx \showDOI      \undefined \def \showDOI       #1{#1}\fi
\ifx \showISBNx    \undefined \def \showISBNx     #1{\unskip}     \fi
\ifx \showISBNxiii \undefined \def \showISBNxiii  #1{\unskip}     \fi
\ifx \showISSN     \undefined \def \showISSN      #1{\unskip}     \fi
\ifx \showLCCN     \undefined \def \showLCCN      #1{\unskip}     \fi
\ifx \shownote     \undefined \def \shownote      #1{#1}          \fi
\ifx \showarticletitle \undefined \def \showarticletitle #1{#1}   \fi
\ifx \showURL      \undefined \def \showURL       {\relax}        \fi
% invisible to TeX
\providecommand\bibfield[2]{#2}
\providecommand\bibinfo[2]{#2}
\providecommand\natexlab[1]{#1}
\providecommand\showeprint[2][]{arXiv:#2}

\bibitem[Abbe(2018)]%
        {abbe2018community}
\bibfield{author}{\bibinfo{person}{Emmanuel Abbe}.} \bibinfo{year}{2018}\natexlab{}.
\newblock \showarticletitle{Community detection and stochastic block models: recent developments}.
\newblock \bibinfo{journal}{\emph{Journal of Machine Learning Research}} \bibinfo{volume}{18}, \bibinfo{number}{177} (\bibinfo{year}{2018}), \bibinfo{pages}{1--86}.
\newblock


\bibitem[Ball et~al\mbox{.}(2011)]%
        {ball2011efficient}
\bibfield{author}{\bibinfo{person}{B. Ball}, \bibinfo{person}{B. Karrer}, {and} \bibinfo{person}{M.~EJ. Newman}.} \bibinfo{year}{2011}\natexlab{}.
\newblock \showarticletitle{Efficient and principled method for detecting communities in networks}.
\newblock \bibinfo{journal}{\emph{Physical Review E}} \bibinfo{volume}{84}, \bibinfo{number}{3} (\bibinfo{year}{2011}), \bibinfo{pages}{036103}.
\newblock


\bibitem[Blondel et~al\mbox{.}(2008)]%
        {com-blondel08}
\bibfield{author}{\bibinfo{person}{V. Blondel}, \bibinfo{person}{J. Guillaume}, \bibinfo{person}{R. Lambiotte}, {and} \bibinfo{person}{E. Lefebvre}.} \bibinfo{year}{2008}\natexlab{}.
\newblock \showarticletitle{{Fast unfolding of communities in large networks}}.
\newblock \bibinfo{journal}{\emph{Journal of Statistical Mechanics: Theory and Experiment}} \bibinfo{volume}{2008}, \bibinfo{number}{10} (\bibinfo{date}{Oct} \bibinfo{year}{2008}), \bibinfo{pages}{P10008}.
\newblock


\bibitem[Brandes et~al\mbox{.}(2007)]%
        {com-brandes07}
\bibfield{author}{\bibinfo{person}{U. Brandes}, \bibinfo{person}{D. Delling}, \bibinfo{person}{M. Gaertler}, \bibinfo{person}{R. Gorke}, \bibinfo{person}{M. Hoefer}, \bibinfo{person}{Z. Nikoloski}, {and} \bibinfo{person}{D. Wagner}.} \bibinfo{year}{2007}\natexlab{}.
\newblock \showarticletitle{{On modularity clustering}}.
\newblock \bibinfo{journal}{\emph{IEEE transactions on knowledge and data engineering}} \bibinfo{volume}{20}, \bibinfo{number}{2} (\bibinfo{year}{2007}), \bibinfo{pages}{172--188}.
\newblock


\bibitem[Chatterjee and Saha(2019)]%
        {com-chatterjee19}
\bibfield{author}{\bibinfo{person}{B. Chatterjee} {and} \bibinfo{person}{H. Saha}.} \bibinfo{year}{2019}\natexlab{}.
\newblock \showarticletitle{Detection of communities in large scale networks}. In \bibinfo{booktitle}{\emph{IEEE 10th Annual Information Technology, Electronics and Mobile Communication Conference (IEMCON)}}. \bibinfo{publisher}{IEEE}, \bibinfo{pages}{1051--1060}.
\newblock
\showISBNx{978-1-7281-2530-5}


\bibitem[Cheong et~al\mbox{.}(2013)]%
        {com-cheong13}
\bibfield{author}{\bibinfo{person}{C. Cheong}, \bibinfo{person}{H. Huynh}, \bibinfo{person}{D. Lo}, {and} \bibinfo{person}{R. Goh}.} \bibinfo{year}{2013}\natexlab{}.
\newblock \showarticletitle{{Hierarchical Parallel Algorithm for Modularity-Based Community Detection Using GPUs}}. In \bibinfo{booktitle}{\emph{Proceedings of the 19th International Conference on Parallel Processing}} (Aachen, Germany) \emph{(\bibinfo{series}{Euro-Par'13})}. \bibinfo{publisher}{Springer-Verlag}, \bibinfo{address}{Berlin, Heidelberg}, \bibinfo{pages}{775--787}.
\newblock
\showISBNx{9783642400469}


\bibitem[Clauset et~al\mbox{.}(2004)]%
        {clauset2004finding}
\bibfield{author}{\bibinfo{person}{Aaron Clauset}, \bibinfo{person}{Mark~EJ Newman}, {and} \bibinfo{person}{Cristopher Moore}.} \bibinfo{year}{2004}\natexlab{}.
\newblock \showarticletitle{Finding community structure in very large networks}.
\newblock \bibinfo{journal}{\emph{Physical review E}} \bibinfo{volume}{70}, \bibinfo{number}{6} (\bibinfo{year}{2004}), \bibinfo{pages}{066111}.
\newblock


\bibitem[Coscia et~al\mbox{.}(2011)]%
        {coscia2011classification}
\bibfield{author}{\bibinfo{person}{Michele Coscia}, \bibinfo{person}{Fosca Giannotti}, {and} \bibinfo{person}{Dino Pedreschi}.} \bibinfo{year}{2011}\natexlab{}.
\newblock \showarticletitle{A classification for community discovery methods in complex networks}.
\newblock \bibinfo{journal}{\emph{Statistical Analysis and Data Mining: The ASA Data Science Journal}} \bibinfo{volume}{4}, \bibinfo{number}{5} (\bibinfo{year}{2011}), \bibinfo{pages}{512--546}.
\newblock


\bibitem[Duch and Arenas(2005)]%
        {duch2005community}
\bibfield{author}{\bibinfo{person}{Jordi Duch} {and} \bibinfo{person}{Alex Arenas}.} \bibinfo{year}{2005}\natexlab{}.
\newblock \showarticletitle{Community detection in complex networks using extremal optimization}.
\newblock \bibinfo{journal}{\emph{Physical review E}} \bibinfo{volume}{72}, \bibinfo{number}{2} (\bibinfo{year}{2005}), \bibinfo{pages}{027104}.
\newblock


\bibitem[Fazlali et~al\mbox{.}(2017)]%
        {com-fazlali17}
\bibfield{author}{\bibinfo{person}{M. Fazlali}, \bibinfo{person}{E. Moradi}, {and} \bibinfo{person}{H. Malazi}.} \bibinfo{year}{2017}\natexlab{}.
\newblock \showarticletitle{{Adaptive parallel Louvain community detection on a multicore platform}}.
\newblock \bibinfo{journal}{\emph{Microprocessors and microsystems}}  \bibinfo{volume}{54} (\bibinfo{date}{Oct} \bibinfo{year}{2017}), \bibinfo{pages}{26--34}.
\newblock


\bibitem[Fortunato(2010)]%
        {com-fortunato10}
\bibfield{author}{\bibinfo{person}{S. Fortunato}.} \bibinfo{year}{2010}\natexlab{}.
\newblock \showarticletitle{{Community detection in graphs}}.
\newblock \bibinfo{journal}{\emph{Physics reports}} \bibinfo{volume}{486}, \bibinfo{number}{3-5} (\bibinfo{year}{2010}), \bibinfo{pages}{75--174}.
\newblock


\bibitem[Ghosh et~al\mbox{.}(2018)]%
        {ghosh2018scalable}
\bibfield{author}{\bibinfo{person}{S. Ghosh}, \bibinfo{person}{M. Halappanavar}, \bibinfo{person}{A. Tumeo}, \bibinfo{person}{A. Kalyanaraman}, {and} \bibinfo{person}{A.H. Gebremedhin}.} \bibinfo{year}{2018}\natexlab{}.
\newblock \showarticletitle{Scalable distributed memory community detection using vite}. In \bibinfo{booktitle}{\emph{2018 IEEE High Performance extreme Computing Conference (HPEC)}}. IEEE, \bibinfo{pages}{1--7}.
\newblock


\bibitem[Gregory(2010)]%
        {com-gregory10}
\bibfield{author}{\bibinfo{person}{S. Gregory}.} \bibinfo{year}{2010}\natexlab{}.
\newblock \showarticletitle{{Finding overlapping communities in networks by label propagation}}.
\newblock \bibinfo{journal}{\emph{New Journal of Physics}}  \bibinfo{volume}{12} (\bibinfo{date}{10} \bibinfo{year}{2010}), \bibinfo{pages}{103018}.
\newblock
Issue 10.


\bibitem[Hafez et~al\mbox{.}(2014)]%
        {hafez2014bnem}
\bibfield{author}{\bibinfo{person}{Ahmed~Ibrahem Hafez}, \bibinfo{person}{Aboul~Ella Hassanien}, {and} \bibinfo{person}{Aly~A Fahmy}.} \bibinfo{year}{2014}\natexlab{}.
\newblock \showarticletitle{BNEM: a fast community detection algorithm using generative models}.
\newblock \bibinfo{journal}{\emph{Social Network Analysis and Mining}}  \bibinfo{volume}{4} (\bibinfo{year}{2014}), \bibinfo{pages}{1--20}.
\newblock


\bibitem[Halappanavar et~al\mbox{.}(2017)]%
        {com-halappanavar17}
\bibfield{author}{\bibinfo{person}{M. Halappanavar}, \bibinfo{person}{H. Lu}, \bibinfo{person}{A. Kalyanaraman}, {and} \bibinfo{person}{A. Tumeo}.} \bibinfo{year}{2017}\natexlab{}.
\newblock \showarticletitle{{Scalable static and dynamic community detection using Grappolo}}. In \bibinfo{booktitle}{\emph{IEEE High Performance Extreme Computing Conference (HPEC)}}. \bibinfo{publisher}{IEEE}, \bibinfo{address}{Waltham, MA USA}, \bibinfo{pages}{1--6}.
\newblock
\showISBNx{978-1-5386-3472-1}


\bibitem[Hesamipour et~al\mbox{.}(2022)]%
        {hesamipour2022detecting}
\bibfield{author}{\bibinfo{person}{Sajjad Hesamipour}, \bibinfo{person}{Mohammad~Ali Balafar}, \bibinfo{person}{Saeed Mousazadeh}, {et~al\mbox{.}}} \bibinfo{year}{2022}\natexlab{}.
\newblock \showarticletitle{Detecting communities in complex networks using an adaptive genetic algorithm and node similarity-based encoding}.
\newblock \bibinfo{journal}{\emph{Complexity}}  \bibinfo{volume}{2023} (\bibinfo{year}{2022}).
\newblock


\bibitem[Kang et~al\mbox{.}(2023)]%
        {kang2023cugraph}
\bibfield{author}{\bibinfo{person}{S. Kang}, \bibinfo{person}{C. Hastings}, \bibinfo{person}{J. Eaton}, {and} \bibinfo{person}{B. Rees}.} \bibinfo{year}{2023}\natexlab{}.
\newblock \showarticletitle{cuGraph C++ primitives: vertex/edge-centric building blocks for parallel graph computing}. In \bibinfo{booktitle}{\emph{IEEE International Parallel and Distributed Processing Symposium Workshops}}. \bibinfo{pages}{226--229}.
\newblock


\bibitem[Kloster and Gleich(2014)]%
        {com-kloster14}
\bibfield{author}{\bibinfo{person}{K. Kloster} {and} \bibinfo{person}{D. Gleich}.} \bibinfo{year}{2014}\natexlab{}.
\newblock \showarticletitle{{Heat kernel based community detection}}. In \bibinfo{booktitle}{\emph{Proceedings of the 20th ACM SIGKDD international conference on Knowledge discovery and data mining}}. \bibinfo{publisher}{ACM}, \bibinfo{address}{New York, USA}, \bibinfo{pages}{1386--1395}.
\newblock


\bibitem[Kolodziej et~al\mbox{.}(2019)]%
        {suite19}
\bibfield{author}{\bibinfo{person}{S. Kolodziej}, \bibinfo{person}{M. Aznaveh}, \bibinfo{person}{M. Bullock}, \bibinfo{person}{J. David}, \bibinfo{person}{T. Davis}, \bibinfo{person}{M. Henderson}, \bibinfo{person}{Y. Hu}, {and} \bibinfo{person}{R. Sandstrom}.} \bibinfo{year}{2019}\natexlab{}.
\newblock \showarticletitle{{The SuiteSparse matrix collection website interface}}.
\newblock \bibinfo{journal}{\emph{JOSS}} \bibinfo{volume}{4}, \bibinfo{number}{35} (\bibinfo{year}{2019}), \bibinfo{pages}{1244}.
\newblock


\bibitem[Lancichinetti and Fortunato(2009)]%
        {com-lancichinetti09}
\bibfield{author}{\bibinfo{person}{A. Lancichinetti} {and} \bibinfo{person}{S. Fortunato}.} \bibinfo{year}{2009}\natexlab{}.
\newblock \showarticletitle{Community detection algorithms: a comparative analysis.}
\newblock \bibinfo{journal}{\emph{Physical Review. E, Statistical, Nonlinear, and Soft Matter Physics}} \bibinfo{volume}{80}, \bibinfo{number}{5 Pt 2} (\bibinfo{date}{Nov} \bibinfo{year}{2009}), \bibinfo{pages}{056117}.
\newblock


\bibitem[Leskovec(2021)]%
        {com-leskovec21}
\bibfield{author}{\bibinfo{person}{J. Leskovec}.} \bibinfo{year}{2021}\natexlab{}.
\newblock \bibinfo{title}{{CS224W: Machine Learning with Graphs | 2021 | Lecture 13.3 - Louvain Algorithm}}.
\newblock
\newblock
\urldef\tempurl%
\url{https://www.youtube.com/watch?v=0zuiLBOIcsw}
\showURL{%
\tempurl}


\bibitem[Lu et~al\mbox{.}(2015)]%
        {com-lu15}
\bibfield{author}{\bibinfo{person}{H. Lu}, \bibinfo{person}{M. Halappanavar}, {and} \bibinfo{person}{A. Kalyanaraman}.} \bibinfo{year}{2015}\natexlab{}.
\newblock \showarticletitle{{Parallel heuristics for scalable community detection}}.
\newblock \bibinfo{journal}{\emph{Parallel computing}}  \bibinfo{volume}{47} (\bibinfo{date}{Aug} \bibinfo{year}{2015}), \bibinfo{pages}{19--37}.
\newblock


\bibitem[Luecken(2016)]%
        {luecken2016application}
\bibfield{author}{\bibinfo{person}{Malte Luecken}.} \bibinfo{year}{2016}\natexlab{}.
\newblock \emph{\bibinfo{title}{Application of multi-resolution partitioning of interaction networks to the study of complex disease}}.
\newblock \bibinfo{thesistype}{Ph.\,D. Dissertation}. \bibinfo{school}{University of Oxford}.
\newblock


\bibitem[Newman(2006)]%
        {newman2006modularity}
\bibfield{author}{\bibinfo{person}{Mark~EJ Newman}.} \bibinfo{year}{2006}\natexlab{}.
\newblock \showarticletitle{Modularity and community structure in networks}.
\newblock \bibinfo{journal}{\emph{Proceedings of the national academy of sciences}} \bibinfo{volume}{103}, \bibinfo{number}{23} (\bibinfo{year}{2006}), \bibinfo{pages}{8577--8582}.
\newblock


\bibitem[Nguyen({[n.\,d.]})]%
        {nguyenleiden}
\bibfield{author}{\bibinfo{person}{Fabian Nguyen}.} \bibinfo{year}{[n.\,d.]}\natexlab{}.
\newblock \emph{\bibinfo{title}{Leiden-Based Parallel Community Detection}}.
\newblock Bachelor's Thesis. \bibinfo{school}{Karlsruhe Institute of Technology, 2021 (zitiert auf S. 31)}.
\newblock


\bibitem[Raghavan et~al\mbox{.}(2007)]%
        {com-raghavan07}
\bibfield{author}{\bibinfo{person}{U. Raghavan}, \bibinfo{person}{R. Albert}, {and} \bibinfo{person}{S. Kumara}.} \bibinfo{year}{2007}\natexlab{}.
\newblock \showarticletitle{{Near linear time algorithm to detect community structures in large-scale networks}}.
\newblock \bibinfo{journal}{\emph{Physical Review E}} \bibinfo{volume}{76}, \bibinfo{number}{3} (\bibinfo{date}{Sep} \bibinfo{year}{2007}), \bibinfo{pages}{036106--1--036106--11}.
\newblock


\bibitem[Reichardt and Bornholdt(2006)]%
        {reichardt2006statistical}
\bibfield{author}{\bibinfo{person}{J{\"o}rg Reichardt} {and} \bibinfo{person}{Stefan Bornholdt}.} \bibinfo{year}{2006}\natexlab{}.
\newblock \showarticletitle{Statistical mechanics of community detection}.
\newblock \bibinfo{journal}{\emph{Physical review E}} \bibinfo{volume}{74}, \bibinfo{number}{1} (\bibinfo{year}{2006}), \bibinfo{pages}{016110}.
\newblock


\bibitem[Rosvall and Bergstrom(2008)]%
        {com-rosvall08}
\bibfield{author}{\bibinfo{person}{M. Rosvall} {and} \bibinfo{person}{C. Bergstrom}.} \bibinfo{year}{2008}\natexlab{}.
\newblock \showarticletitle{{Maps of random walks on complex networks reveal community structure}}.
\newblock \bibinfo{journal}{\emph{Proceedings of the national academy of sciences}} \bibinfo{volume}{105}, \bibinfo{number}{4} (\bibinfo{year}{2008}), \bibinfo{pages}{1118--1123}.
\newblock


\bibitem[Rotta and Noack(2011)]%
        {com-rotta11}
\bibfield{author}{\bibinfo{person}{R. Rotta} {and} \bibinfo{person}{A. Noack}.} \bibinfo{year}{2011}\natexlab{}.
\newblock \showarticletitle{Multilevel local search algorithms for modularity clustering}.
\newblock \bibinfo{journal}{\emph{Journal of Experimental Algorithmics (JEA)}}  \bibinfo{volume}{16} (\bibinfo{year}{2011}), \bibinfo{pages}{2--1}.
\newblock


\bibitem[Ryu and Kim(2016)]%
        {com-ryu16}
\bibfield{author}{\bibinfo{person}{S. Ryu} {and} \bibinfo{person}{D. Kim}.} \bibinfo{year}{2016}\natexlab{}.
\newblock \showarticletitle{{Quick community detection of big graph data using modified louvain algorithm}}. In \bibinfo{booktitle}{\emph{IEEE 18th International Conference on High Performance Computing and Communications (HPCC)}}. \bibinfo{publisher}{IEEE}, \bibinfo{address}{Sydney, NSW}, \bibinfo{pages}{1442--1445}.
\newblock
\showISBNx{978-1-5090-4297-5}


\bibitem[Sahu(2023a)]%
        {sahu2023gveleiden}
\bibfield{author}{\bibinfo{person}{Subhajit Sahu}.} \bibinfo{year}{2023}\natexlab{a}.
\newblock \showarticletitle{GVE-Leiden: Fast Leiden Algorithm for Community Detection in Shared Memory Setting}.
\newblock \bibinfo{journal}{\emph{arXiv preprint arXiv:2312.13936}} (\bibinfo{year}{2023}).
\newblock


\bibitem[Sahu(2023b)]%
        {sahu2023gvelouvain}
\bibfield{author}{\bibinfo{person}{Subhajit Sahu}.} \bibinfo{year}{2023}\natexlab{b}.
\newblock \showarticletitle{GVE-Louvain: Fast Louvain Algorithm for Community Detection in Shared Memory Setting}.
\newblock \bibinfo{journal}{\emph{arXiv preprint arXiv:2312.04876}} (\bibinfo{year}{2023}).
\newblock


\bibitem[Shi et~al\mbox{.}(2021)]%
        {com-shi21}
\bibfield{author}{\bibinfo{person}{J. Shi}, \bibinfo{person}{L. Dhulipala}, \bibinfo{person}{D. Eisenstat}, \bibinfo{person}{J. {\L}{\k{a}}cki}, {and} \bibinfo{person}{V. Mirrokni}.} \bibinfo{year}{2021}\natexlab{}.
\newblock \bibinfo{title}{Scalable community detection via parallel correlation clustering}.
\newblock
\newblock


\bibitem[Souravlas et~al\mbox{.}(2021)]%
        {com-souravlas21}
\bibfield{author}{\bibinfo{person}{S. Souravlas}, \bibinfo{person}{A. Sifaleras}, \bibinfo{person}{M. Tsintogianni}, {and} \bibinfo{person}{S. Katsavounis}.} \bibinfo{year}{2021}\natexlab{}.
\newblock \showarticletitle{A classification of community detection methods in social networks: a survey}.
\newblock \bibinfo{journal}{\emph{International journal of general systems}} \bibinfo{volume}{50}, \bibinfo{number}{1} (\bibinfo{date}{Jan} \bibinfo{year}{2021}), \bibinfo{pages}{63--91}.
\newblock


\bibitem[Staudt et~al\mbox{.}(2016)]%
        {staudt2016networkit}
\bibfield{author}{\bibinfo{person}{C.L. Staudt}, \bibinfo{person}{A. Sazonovs}, {and} \bibinfo{person}{H. Meyerhenke}.} \bibinfo{year}{2016}\natexlab{}.
\newblock \showarticletitle{NetworKit: A tool suite for large-scale complex network analysis}.
\newblock \bibinfo{journal}{\emph{Network Science}} \bibinfo{volume}{4}, \bibinfo{number}{4} (\bibinfo{year}{2016}), \bibinfo{pages}{508--530}.
\newblock


\bibitem[Traag(2024)]%
        {traag2024leiden}
\bibfield{author}{\bibinfo{person}{V. Traag}.} \bibinfo{year}{2024}\natexlab{}.
\newblock \showarticletitle{Personal communication}.
\newblock  (\bibinfo{year}{2024}).
\newblock


\bibitem[Traag and {\v{S}}ubelj(2023)]%
        {traag2023large}
\bibfield{author}{\bibinfo{person}{V.A. Traag} {and} \bibinfo{person}{L. {\v{S}}ubelj}.} \bibinfo{year}{2023}\natexlab{}.
\newblock \showarticletitle{Large network community detection by fast label propagation}.
\newblock \bibinfo{journal}{\emph{Scientific Reports}} \bibinfo{volume}{13}, \bibinfo{number}{1} (\bibinfo{year}{2023}), \bibinfo{pages}{2701}.
\newblock


\bibitem[Traag et~al\mbox{.}(2019)]%
        {com-traag19}
\bibfield{author}{\bibinfo{person}{V. Traag}, \bibinfo{person}{L. Waltman}, {and} \bibinfo{person}{N. Eck}.} \bibinfo{year}{2019}\natexlab{}.
\newblock \showarticletitle{{From Louvain to Leiden: guaranteeing well-connected communities.}}
\newblock \bibinfo{journal}{\emph{Scientific Reports}} \bibinfo{volume}{9}, \bibinfo{number}{1} (\bibinfo{date}{Mar} \bibinfo{year}{2019}), \bibinfo{pages}{5233}.
\newblock


\bibitem[Waltman and Eck(2013)]%
        {com-waltman13}
\bibfield{author}{\bibinfo{person}{L. Waltman} {and} \bibinfo{person}{N. Eck}.} \bibinfo{year}{2013}\natexlab{}.
\newblock \showarticletitle{A smart local moving algorithm for large-scale modularity-based community detection}.
\newblock \bibinfo{journal}{\emph{The European physical journal B}} \bibinfo{volume}{86}, \bibinfo{number}{11} (\bibinfo{year}{2013}), \bibinfo{pages}{1--14}.
\newblock


\bibitem[Whang et~al\mbox{.}(2013)]%
        {com-whang13}
\bibfield{author}{\bibinfo{person}{J. Whang}, \bibinfo{person}{D. Gleich}, {and} \bibinfo{person}{I. Dhillon}.} \bibinfo{year}{2013}\natexlab{}.
\newblock \showarticletitle{{Overlapping community detection using seed set expansion}}. In \bibinfo{booktitle}{\emph{Proceedings of the 22nd ACM international conference on Information \& Knowledge Management}}. \bibinfo{pages}{2099--2108}.
\newblock


\bibitem[Wickramaarachchi et~al\mbox{.}(2014)]%
        {com-wickramaarachchi14}
\bibfield{author}{\bibinfo{person}{C. Wickramaarachchi}, \bibinfo{person}{M. Frincu}, \bibinfo{person}{P. Small}, {and} \bibinfo{person}{V. Prasanna}.} \bibinfo{year}{2014}\natexlab{}.
\newblock \showarticletitle{Fast parallel algorithm for unfolding of communities in large graphs}. In \bibinfo{booktitle}{\emph{IEEE High Performance Extreme Computing Conference (HPEC)}}. IEEE, \bibinfo{publisher}{IEEE}, \bibinfo{address}{Waltham, MA USA}, \bibinfo{pages}{1--6}.
\newblock


\bibitem[Wolf et~al\mbox{.}(2019)]%
        {wolf2019paga}
\bibfield{author}{\bibinfo{person}{F~Alexander Wolf}, \bibinfo{person}{Fiona~K Hamey}, \bibinfo{person}{Mireya Plass}, \bibinfo{person}{Jordi Solana}, \bibinfo{person}{Joakim~S Dahlin}, \bibinfo{person}{Berthold G{\"o}ttgens}, \bibinfo{person}{Nikolaus Rajewsky}, \bibinfo{person}{Lukas Simon}, {and} \bibinfo{person}{Fabian~J Theis}.} \bibinfo{year}{2019}\natexlab{}.
\newblock \showarticletitle{PAGA: graph abstraction reconciles clustering with trajectory inference through a topology preserving map of single cells}.
\newblock \bibinfo{journal}{\emph{Genome biology}}  \bibinfo{volume}{20} (\bibinfo{year}{2019}), \bibinfo{pages}{1--9}.
\newblock


\bibitem[Xie et~al\mbox{.}(2011)]%
        {com-xie11}
\bibfield{author}{\bibinfo{person}{J. Xie}, \bibinfo{person}{B. Szymanski}, {and} \bibinfo{person}{X. Liu}.} \bibinfo{year}{2011}\natexlab{}.
\newblock \showarticletitle{{SLPA: Uncovering overlapping communities in social networks via a speaker-listener interaction dynamic process}}. In \bibinfo{booktitle}{\emph{IEEE 11th International Conference on Data Mining Workshops}}. IEEE, \bibinfo{publisher}{IEEE}, \bibinfo{address}{Vancouver, Canada}, \bibinfo{pages}{344--349}.
\newblock


\bibitem[Yang et~al\mbox{.}(2016)]%
        {yang2016comparative}
\bibfield{author}{\bibinfo{person}{Zhao Yang}, \bibinfo{person}{Ren{\'e} Algesheimer}, {and} \bibinfo{person}{Claudio~J Tessone}.} \bibinfo{year}{2016}\natexlab{}.
\newblock \showarticletitle{A comparative analysis of community detection algorithms on artificial networks}.
\newblock \bibinfo{journal}{\emph{Scientific reports}} \bibinfo{volume}{6}, \bibinfo{number}{1} (\bibinfo{year}{2016}), \bibinfo{pages}{30750}.
\newblock


\bibitem[You et~al\mbox{.}(2020)]%
        {com-you20}
\bibfield{author}{\bibinfo{person}{X. You}, \bibinfo{person}{Y. Ma}, {and} \bibinfo{person}{Z. Liu}.} \bibinfo{year}{2020}\natexlab{}.
\newblock \showarticletitle{{A three-stage algorithm on community detection in social networks}}.
\newblock \bibinfo{journal}{\emph{Knowledge-Based Systems}}  \bibinfo{volume}{187} (\bibinfo{year}{2020}), \bibinfo{pages}{104822}.
\newblock


\bibitem[Zarayeneh and Kalyanaraman(2021)]%
        {com-zarayeneh21}
\bibfield{author}{\bibinfo{person}{N. Zarayeneh} {and} \bibinfo{person}{A. Kalyanaraman}.} \bibinfo{year}{2021}\natexlab{}.
\newblock \showarticletitle{{Delta-Screening: A Fast and Efficient Technique to Update Communities in Dynamic Graphs}}.
\newblock \bibinfo{journal}{\emph{IEEE transactions on network science and engineering}} \bibinfo{volume}{8}, \bibinfo{number}{2} (\bibinfo{date}{Apr} \bibinfo{year}{2021}), \bibinfo{pages}{1614--1629}.
\newblock


\bibitem[Zeng and Yu(2015)]%
        {com-zeng15}
\bibfield{author}{\bibinfo{person}{J. Zeng} {and} \bibinfo{person}{H. Yu}.} \bibinfo{year}{2015}\natexlab{}.
\newblock \showarticletitle{{Parallel Modularity-Based Community Detection on Large-Scale Graphs}}. In \bibinfo{booktitle}{\emph{IEEE International Conference on Cluster Computing}}. \bibinfo{pages}{1--10}.
\newblock
\showISBNx{978-1-4673-6598-7}


\end{thebibliography}
